\documentclass{article}

\input{tcilatex}
\begin{document}

\bigskip \baselineskip1cm \textwidth16.5 cm

\begin{center}
\textbf{The second order dense ferromagnetic-ferromagnetic phase transition }

Aycan \"{O}zkan, B\"{u}lent Kutlu

Gazi \"{U}niversitesi, Fen -Edebiyat Fak\"{u}ltesi, Fizik B\"{o}l\"{u}m\"{u}
, 06500 Teknikokullar, Ankara, Turkey,

e-mail: aycan@gazi.edu.tr, bkutlu@gazi.edu.tr
\end{center}

The fcc spin-1 Ising (BEG) model has a dense ferromagnetic ($df$) ground
state instead of the ferromagnetic ground state at low temperature region
and exhibits the dense ferromagnetic ($df$) - ferromagnetic ($F$) phase
transition for $d=D/J=2.9$, $k=K/J=-0.5$, $\ell =L/J=0$ and $h=H/J=0$. The
critical behavior of the dense ferromagnetic ($df$) - ferromagnetic ($F$)
phase transition has been investigated using the cellular automaton cooling
and heating algorithms. The universality class and the type of the dense
ferromagnetic ($df$) - ferromagnetic ($F$) phase transition have been
researched within the framework of the finite - size scaling, the power law
relations and the probability distribution. The results show that the dense
ferromagnetic- ferromagnetic phase transition is of the second order and the
model shows universal second order Ising critical behavior at $d=2.9$
parameter value through $k=-0.5$ line.

PACS number(s): 05.10.-a, 05.50.+q, 64.60.-i

\textbf{I . INTRODUCTION}

In recent years, some of the studies indicated that the spin-1 Ising model
has a ground state ordered structure which is named the dense ferromagnetic
( $df$) $\left[ 1-4\right] $. In absence of the $df$ ordered structure, the
phase diagrams were formed for some phase transitions which has been assumed
the weak first order transition instead of the second order transition. The
presence of the $df$ ordered structure can be caused to some changes on the
global phase diagrams $\left[ 4\right] $. This case clarifies the
differences among the results of the previous studies for the ($kT_{C}/J$, $%
d $) phase diagram through the $k$ line. $[5-11]$. For example, the phase
diagram has exhibited a tricritical point ($TCP$) instead of a critical end
point ($CEP$) for $k=-0.5$. While the MFA $\left[ 7\right] $ and RG $\left[ 8%
\right] $\ studies exhibited a critical end point ($CEP$) for k=-0.5 on the
BEG model global phase diagram, CA $[5,6]$, MCRG $[9],$ TPCA $\left[ 10%
\right] $ and CVM $[11]$ studies show that there is the tricritical point ($%
TCP$). Through the $k$ line, the $df$ - $F$ phase transition is very
important for determining the type of the phase boundary. The purpose of
this study is to define the $df$ ordered phase and is to investigate the
nature of $df-F$ phase transition at $d=2.9$ value through $k=-0.5$ \ line.
This point creates the type of the special point of the ( $kT_{C}/J$ , $d$ )
phase space for $k=-0.5$. Therefore, we have found the $df$ - $F$ phase
transition worthy of investigation in depth. Furthermore, the universality
class of the $df$ - $F$ phase transition has not been investigated so far.
The critical temperature and the statical critical exponents are estimated
by analyzing the data within the framework of the finite - size scaling
theory and the power law relations.

The spin-1 Ising model, which is known as the generalized
Blume-Emery-Griffiths (BEG) model, can be used to simulate many physical
systems. The model firstly has been presented for describing phase
separation and superfluid ordering in He mixtures $\left[ 12\right] $. The
versions of the model have been applied to the physical systems such as the
solid-liquid-gas systems $\left[ 13\right] $, the multicomponent fluids $%
\left[ 14\right] $, the microemulsions $\left[ 15\right] $, the
semiconductor alloys $\left[ 16-18\right] $, He$^{3}$-He$^{4}$ mixtures $%
\left[ 12,19\right] $ and the binary alloys $\left[ 20\right] $.\ 

The BEG model Hamiltonian is defined as\qquad

\begin{equation}
H_{I}=-J\sum_{<ij>}S_{i}S_{j}-K\sum_{<ij>}S_{i}^{2}S_{j}^{2}+L%
\sum_{<ij>}(S_{i}^{2}S_{j}+S_{i}S_{j}^{2})+D\sum_{i}S_{i}^{2}+h\sum_{i}S_{i}
\end{equation}%
which is equivalent to the lattice gas Hamiltonian under some
transformations $\left[ 21-23\right] $. $\left\langle ij\right\rangle $
denotes summation over all nearest-neighbor (nn) pairs of sites and $%
S_{i}=-1,$ $0,$ $1$. The parameters $J$, $K$, $L$, $D$ and $h$ are bilinear,
biquadratic, dipole-quadrupole interaction terms, the single-ion anisotropy
constant and the field term. The BEG model for $k\geq 0$ has been studied by
mean field approximation \ (MFA) $\left[ 12-14\right] $, the transfer matrix
method $\left[ 24\right] $, series expansion method $\left[ 25\right] $, the
constant coupling approximation $\left[ 26\right] $, the position-space
renormalization method $\left[ 27\right] $, cluster variation method (CVM) $%
\left[ 1\right] $, linear-chain approximation $\left[ 2\right] $, Monte
Carlo method (MC) $\left[ 3\right] $ and Cellular Automaton (CA) $\left[
4-6,28,29\right] $.

In this paper, the fcc BEG model for $d=D/J=2.9$, $k=K/J=-0.5$, $\ell =L/J=0$
and $h=H/J=0$ is simulated using cooling and heating algorithm improved from
Creutz Cellular Automaton. In the previous papers, the Creutz cellular
automaton (CCA) algorithm and its improved versions have been used
successfully to study the properties of the critical behaviors of the Ising
model Hamiltonians $\left[ 4-6,28-50\right] $. The CCA algorithm, which was
first introduced by Creutz $\left[ 51\right] $, is a microcanonical
algorithm interpolating between the conventional Monte Carlo and the
molecular dynamics techniques. The Creutz cellular automaton (CCA) is faster
than the conventional Monte Carlo method (MC). The CCA does not need high
quality random numbers and it is a new and an alternative simulation method
for physical systems. It has another advantage allowing the specific heat to
be computed from internal energy fluctuations. Our previous studies showed
that the heating and the cooling algorithms improved from the Creutz
Cellular Automaton algorithm are effective to study the phase space and the
critical behavior of the Blume Emery Griffiths model $\left[ 4-6,28-30,35,36%
\right] $.

\textbf{II . RESULTS AND DISCUSSION}

The CA algorithm of spin-1 Ising model is a microcanonical algorithm. The
total energy $H$, which is conserved, is given by

\begin{equation}
H=H_{I}+H_{K}
\end{equation}%
where $H_{I}$ is Ising energy which is given by equation1 and $H_{K}$ is
kinetic energy.

The kinetic energy $H_{K}$ is an integer, equal to the change in the Ising
spin energy for any spin flip and its value lie in the interval ($0$, $m$). $%
m$ is equal to $24J$ for $d=D/J=2.9$, $k=K/J=-0.5$, $\ell =L/J=0$ and $%
h=H/J=0$ on fcc lattice. For a site to be updated, its spin is changed to
one of the other two states with $1/2$ probability. If this energy is
transferable to or from the kinetic energy variable of the site, such that
the total energy H is conserved, then this change is done and kinetic energy
is appropriately changed. Otherwise the spin is not change $\left[ 30,34-36%
\right] $.

At the heating and the cooling algorithms, the simulation consist of two
parts, the initialization procedure and the computation of the thermodynamic
quantities. The initial configuration for heating and cooling algorithms can
be set in different shapes. In this study, the initial configurations are
obtained at three different shapes for heating algorithm during 20.000 CA
steps. Firstly, all the spins are up ($S=+1$) at the absolute zero
temperature for both algorithms. The initial configuration of the heating
algorithm has been obtained at low temperature ordered phase ($df$) adding
kinetic energy which is equal to the maximum change in the Ising spin energy
for the any spin flip to the spin system for set I and II. The another
initial configuration (set III) has been obtained flipping $8\%$ of the
spins to $S=0$ state. The heating rate is realized by increasing of $8\%$ in
the kinetic energy ($H_{k}$) of $15\%$ of the fcc lattice for two sets at
the computation of the thermodynamic quantities. At set III, the heating
rate is realized by increasing of $8\%$ in the kinetic energy ($H_{k}$) of
all site of the fcc lattice.

On the other hand, the initial configuration for the cooling algorithm is
obtained adding energy to the $70\%$ of the spin system for getting the
disordered phase ($P$) at high temperature. During the cooling cycle, the
cooling rate is realized by decreasing of $8\%$ in the kinetic energy ($%
H_{k} $) from $25\%$ of the spin system. The initial configurations are run
during the 20.000 Cellular Automaton time steps. Instead of resetting the
starting configuration at each energy, it was used the final configuration
at a given energy as the starting point for the next at both heating and
cooling algorithms. The computed values of the thermodynamic quantities (the
order parameters ($M$, $Q$), the susceptibility ($\chi $), the Ising energy (%
$H_{I} $) and the specific heat ($C$)) are averages over the lattice and
over the number of time steps ($2.000.000$ ) with discard of the first $%
100.000$ time steps during the cellular automaton develops $\left[ 4-6,28,29%
\right] $.

They have been computed on the fcc lattice with $L=$ $8$, $9$, $10$, $11$
and $12$ (The total number of sites is $N=4L^{3}$) for periodic boundary
conditions. The fcc lattice was formed in a simple cubic (sc) lattice. (The
total number of sites is $N=4L^{3}=6912$ for $L=12$ fcc lattice, this total
site number equals to $L=19$ for the simple cubic lattice). The presented
figures are set III (heating algorithm) results.

The order parameters, the Ising energy, susceptibility and specific heat are
calculated from

\begin{equation}
M=\frac{1}{N}\sum_{i}S_{i},\text{ \ \ \ \ \ }Q=\frac{1}{N}\sum_{i}S_{i}^{2}
\end{equation}%
\begin{equation}
U_{I}=(-J\sum_{<ij>}S_{i}S_{j}-K\sum_{<ij>}S_{i}^{2}S_{j}^{2}+D%
\sum_{i}S_{i}^{2})/U_{0}
\end{equation}

\begin{equation}
\chi =N\frac{\left\langle M^{2}\right\rangle -\left\langle M\right\rangle
^{2}}{kT}
\end{equation}

\begin{equation}
C_{I}/k=N\frac{\left\langle U_{I}^{2}\right\rangle -\left\langle
U_{I}\right\rangle ^{2}}{(kT)^{2}}
\end{equation}%
where $U_{0}$ is the ground state energy at $kT/J=0$.

The ferromagnetic ($F$) and the paramagnetic ($P$) phases can be determined
with the average occupation of the states $\left\langle P_{\pm
1,0}\right\rangle $. As the projectors for the states $S=+1$, $-1$ and $0$
are $P_{+1}=\frac{1}{2}S(S+1)$, $P_{-1}=\frac{1}{2}S(S-1)$ and $%
P_{0}=1-S^{2} $, the average occupation of the states are $\left\langle
P_{+1}\right\rangle =\frac{1}{2}(Q+M)$, $\left\langle P_{-1}\right\rangle =%
\frac{1}{2}(Q-M)$ and $\left\langle P_{0}\right\rangle =1-Q$, respectively.
With considering the average occupation of the states, another ferromagnetic
phase can be determined as the dense ferromagnetic phase ($df$).\ 

Ferromagnetic ($F$): $\left\langle P_{+1}\right\rangle \neq \left\langle
P_{-1}\right\rangle \neq \left\langle P_{0}\right\rangle \neq 0$, $(M\neq
Q\neq 0)$

Dense ferromagnetic ($df$): $\left\langle P_{-1}\right\rangle \rightarrow 0$%
; $\left\langle P_{+1}\right\rangle \neq \left\langle P_{0}\right\rangle
\neq 0$, $(M\cong Q\neq 0)$

Paramagnetic ($P$): $\left\langle P_{-1}\right\rangle =\left\langle
P_{+1}\right\rangle \neq \left\langle P_{0}\right\rangle \neq 0$, $%
(M=0,Q\neq 0)$.

\textbf{II .1 Temperature Variations of Thermodynamic Quantities for the }$%
dF-F-P$\textbf{\ Phase Transitions }

The temperature variation of the order parameters ($M$, $Q$), the
susceptibility ($\chi $), the Ising energy ($H_{I}$) and the specific heat ($%
C_{I}$) are illustrated in figure 1 for exhibiting the general aspect of the
successive $df$ $-$ $F$ $-$ $P$ phase transitions at $d=2.9$ parameter value
through $k=-0.5$ line. As it is seen in figure 1(a) and figure 1(c), the
order parameters and the Ising energy appear continuously for $df$ $-F$ and $%
F-P$ phase transitions. Therefore both phase transitions are of the second
order as functional behavior. The susceptibility ($\chi $) and the specific
heat ($C_{I}$, $C$)exhibit two peaks at $T_{C1}$ and $T_{C2}$ temperatures
corresponding to $df-F$ and $F-P$ phase transitions (Figure 1(b), 1(d) and
figure 2). It can be seen from figure 1(c), the functional change of the
Ising energy from order to order ($df$ $-F$) phase transition is different
from the view of the order to disorder ($F-P$) phase transition. For $df$ - $%
F$ phase transition, the Ising energy difference ($\Delta U$) is greater
than for the $F$ - $P$ phase transition. The estimated critical temperature
from susceptibility and specific heat maxima is compatible with each other
for $F$ - $P$ phase transition. But the critical temperature values for $df$
- $F$ phase transition are not compatible (Figure 1(b) and 2(a)). Therefore
specific heat ($C$) has been recalculated for only spin-spin interaction
energy (Figure 2(b)).

The spin-spin interaction energy $U$ is determined as

\begin{equation}
U=(-J\sum_{<ij>}S_{i}S_{j})/U_{0}
\end{equation}%
The specific heat calculated from $U$ can show a sharp peak for $df-F$ phase
transition. Because, the first sum ($U$) in Ising energy ($U_{I}$)
distinguish the $S=+1$ and $-1$ states. Indeed, the obtained infinite
critical temperature ($T_{C1}$($\infty $)$=1.52\pm 0.04$) from the
susceptibility ($\chi $) and specific heat ($C$) peak temperatures are
compatible with each other. $T_{C2}$($\infty $) is obtained from the
susceptibility ($\chi $) and specific heat ($C_{I}$ and $C$) peak
temperatures as $3.20\pm 0.02$.

The temperature variations of the $\left\langle P_{+1}\right\rangle $, $%
\left\langle P_{-1}\right\rangle $ and $\left\langle P_{0}\right\rangle $
are given in figure 3. The initial configuration for the heating algorithm
is created as all spins are up ($S=+1$) at absolute zero temperature ($T=0$%
). If the enough energy is added to the spin system, the $S=0$ begins to
arise. The excitation energy of the single spin flipping from $S=+1$ to $S=0$
is $3.1J$ while it is $24J$ for the single spin flipping from $S=+1$ to $S=-1
$. Therefore, at low temperature region, the rising probability of the $S=-1$
state has to be lower than $S=0$ state. It can be seen in figure 3 that the
spin system includes $S=+1$ and $S=0$ states predominantly. The value of $%
\left\langle P_{+1}\right\rangle $ is about\ 1 and $\left\langle
P_{0}\right\rangle $ is different from zero, while $\left\langle
P_{-1}\right\rangle $ appears almost zero indicating the $df$ phase for$\
T<T_{C1}(L)$. At the same time,. So, $M$ is almost equal to the $Q$ ($M\cong
Q\neq 0$). As $\left\langle P_{-1}\right\rangle $ increases above $T_{C1}(L)$%
, the $df$ ordered phase changes to the $F$ ordered phase. At high
temperature region, the ferromagnetic - the paramagnetic phase transition ($F
$ - $P$) occurs . $\left\langle P_{+1}\right\rangle $ is equal to $%
\left\langle P_{-1}\right\rangle $ and\ the system is in the paramagnetic ($P
$) disordered phase above $T_{C2}$($L$) temperature ( $M\neq Q\neq 0$).
Therefore the phase space is divided into three regions ( $df$, $F$ and $P$
). It is obvious that to follow the temperature variation of $\left\langle
P_{-1}\right\rangle $ is a useful way to prove the existence of the $df$
ordered phase.

\textbf{II .2 Probability Distribution of Order Parameter }

The another useful procedure to distinguish the phase transition type is to
calculate the probability distributions of the order parameter ($P(M)$). In
our study the probability distribution is calculated by

\begin{equation}
P_{L}(M)=\frac{N_{M}}{N_{CCAS}}
\end{equation}
where $N_{M}$ is the number of times that magnetization $M$ appears, and $%
N_{CCAS}$ is the total number of the cellular automaton steps. The histogram
with $200$ bins are used for plotting the probability distribution of the
magnetization $\left[ 24\text{, }31\right] $. The probability distribution
of the order parameter ($P(M)$) near the phase transition temperature shows
two peaks in the second phase transitions.

The probability distributions of the order parameter ($P(M)$) are shown for
different temperature values in figure 4. The peaks of the order parameter
probability distribution exhibits minimum with increasing temperature at the
low temperature region. This minimum corresponds to the second order $df$ $-$
$F$ phase transition at the $T_{C1}^{\chi }(L=12)=1.479$. Although the phase
transition is of the second order, the probability distribution shows the
single peak near the phase transition temperature $T_{C1}(L)$. Because the
system has $S=+1$ and $0$ spins below $T_{C1}(L)$ and the transition is from
order ($df$) to order ($F$). However, the probability distributions in the $F
$ $-$ $P$ phase transition region exhibit the two peaks with the
contribution of the $S=-1$ state near the $T_{C2}^{{}}(L=12)=3.187$. For $%
T>T_{C2}$, there is a single peak focused to $M=0$ indicating the disordered
($P$) phase.

\textbf{II.3 Finite - Size Scaling Analyses and the Statical Critical
Exponents}

The values of the statical critical exponents ($\nu $, $\beta $, $\gamma $, $%
\alpha $) are estimated within the framework of the finite - size scaling
theory and the power laws. The infinite lattice critical temperature $%
T_{C}(\infty )$ has been obtained from the susceptibility and the specific
heat peak temperatures for the successive second order $df$ $-F$ $-$ $P$
phase transitions and from the intersection point of Binder cumulant curves (%
$U_{L}$) for the second order $F$ $-P$ phase transition $\left[ 52\right] $.
The finite - size scaling relations of the Binder cumulant ($U_{L}$), the
order parameter ($M$ ), the susceptibility ($\chi $) and the specific heat ($%
C$) are given by

\begin{equation}
U_{L}=G^{\circ }(\varepsilon L^{1/\nu })
\end{equation}

\begin{equation}
M=L^{-\beta /\nu }X^{\circ }(\varepsilon L^{1/\nu })
\end{equation}

\begin{equation}
kT\chi =L^{\gamma /\nu }Y^{\circ }(\varepsilon L^{1/\nu })
\end{equation}

\begin{equation}
C=L^{a/\nu }Z^{\circ }(\varepsilon L^{1/\nu })
\end{equation}

For large $x=\varepsilon L^{1/\nu }$, the finite lattice critical behaviors
must be asymptotically reproduced, that is,

\begin{equation}
X^{\circ }(x)\propto Ax^{\beta }
\end{equation}

\begin{equation}
Y^{\circ }(x)\propto Bx^{-\gamma }
\end{equation}

\begin{equation}
Z^{\circ }(x)\propto Cx^{-\alpha }
\end{equation}

According to the finite size scaling theory, the data for the finite - size
lattices of the thermodynamic quantities should lie on a single curve for
the temperatures both below and above $T_{C}(\infty )$ with universal
critical exponents. The critical exponents $\beta $, $\gamma $ and $\alpha $
have been obtained from the Log-Log plots of the asymptomatic functions.

The temperature variation of the Binder cumulant is shown for the different
lattice sizes in figure 5. The Binder cumulant curves intersect at the $%
T_{C2}^{U_{L}}(\infty )=3.20\pm 0.02$ corresponding to the $F-P$ phase
transition (Figure 5(a)). This value is compatible with $T_{C2}^{{}}(\infty )
$ which is extrapolated according to the finite size scaling theory from the
susceptibility and the specific heat peak temperatures ($T_{C}(L)$) of the
finite lattices, respectively.

\begin{equation}
T_{C}(L)=T_{C}(\infty )+aL^{-1/\nu }
\end{equation}%
It can be seen in the inset of the figure 5(a) that there is no intersection
at the Binder cumulant for the data of the $df-F$ phase transition region.
However, the Binder cumulant curves exhibit a plateau near the infinite
lattice critical temperature ($T_{C1}(\infty )$) which is obtained from the
susceptibility ($\chi $) and the specific heat ($C$) peak temperatures as $%
T_{C1}^{\chi }(\infty )=1.52\pm 0.04$. In figure 5(b), the scaling data of
the Binder cumulants are shown for the second order phase transition from $F$
to $P$. Near the $T_{C2}^{U_{L}}$($\infty $), Binder cumulant curves have
been scaled well for $T_{C}$=$T_{C2}$($\infty $) with $\nu =0.64$. It can be
seen in the inset of the figure 5(b) that the data corresponding to $df$
ordered phase could not be scaled with the $T_{C2}^{U_{L}}(\infty )$
critical temperature value. However, the scaling data of the binder cumulant
corresponding to the $df$ ordered phase lie on a single curve for $%
\varepsilon =$($T-T_{C1}^{{}}(\infty )$)$/T_{C1}^{{}}(\infty )$ at\text{ }$%
T<T_{C1}^{{}}(\infty )$ region using $\nu =0.64$ (Figure 5(c)) and the
finite size scaling relations validate for the $df$ ordered region.

In figure 6(a), the scaling data of the order parameter is illustrated at
the successive $df-F$ $-$ $P$ phase transitions for $L=8,9,10,11$ and $12$
at $T_{C}^{{}}=T_{C2}^{{}}(\infty )$. The order parameter data lie on the
two different curves with slope=$\beta /\nu =0.31$ and $\beta ^{\prime }/\nu
=-0.55$ for the temperatures both below and above $T_{C2}^{{}}(\infty )$
respectively except for $df$ ordered region with $\beta =0.31$ and $\nu =0.64
$. As it is seen in the inset of the figure 6(a), the data of $df-F$ phase
transition region have not been scaled with $T_{C}=T_{C2}^{{}}(\infty )$.
However, the data corresponding to the $df\ -F$ phase transition region have
been scaled well with $T_{C}=T_{C1}^{{}}(\infty )$ for $T<T_{C1}^{{}}(\infty
)$ using $\beta =0.31$ and $\nu =0.64$ in figure 6(b).

The scaling data of the susceptibility have been shown in figure 7 with the
straight lines describing the theoretically predicted behavior for large $x$
(Equation 8). The susceptibility data for the temperatures both below and
above $T_{C2}^{{}}$($\infty $) agrees with the asymptotic form except for
the $df$ - $F$ phase transition region and so with the $T_{C}=T_{C2}^{{}}(%
\infty )$, $\gamma =\gamma ^{\prime }=1.25$ and $\nu =0.64$ in figure 7(a)
and 7(c). However, the data of $df$ ordered phase ($T<T_{C1}$) have been
scaled with the $T_{C}=T_{C1}^{{}}(\infty )$ using $\gamma =1.25$ and $\nu
=0.64$ in figure 7(b).

The finite size scaling data of the singular portion of the specific heat ($C
$) have been exhibited in figure 8. The data of $F$ $-$ $P$ phase transition
region of $C_{I}$ have been scaled well both below and above $T_{C2}^{C}$($%
\infty $) using $\alpha =0.12$, $\nu =0.64$ and the correction terms, $%
b^{-}=-70$ and $b^{+}=-8$ (Figure 8(a) and 8(b)). Although the data of the $%
df$ phase region could not been scaled with $T_{C}=T_{C2}^{C}(\infty )$ in
figure 8(a), the data lie on single curves with slope=$-\alpha /\nu =-0.12$
at the both side of the $T_{C1}^{C}(\infty )$, using $\alpha =0.12$, $\nu
=0.64$ and the correction terms, $b^{-}=-1$ and $b^{+}=-0.3$ in figure 8(c)
and (d). On the other hand, the specific heat ($C_{I}$) data calculated from 
$U_{I}$ scales well at $df$ - $F$- $P$ phase transitions using $\alpha =0.12$%
, $\nu =0.64$ for $T<T_{C1}^{\chi }(\infty )$and $T<T_{C2}^{C_{I}}(\infty )$
and $T>T_{C2}^{C_{I}}(\infty )$.

The $M$, $\chi $ and $C$ data have been analyzed within the framework of the
finite size scaling theory for the successive $df-F-P$ phase transitions.
The estimated values of the statical critical exponents are in good
agreement with the universal values ($\alpha =0.12$, $\beta =0.31$, $\gamma
=1.25$, $\nu =0.64$) for the $df-F$ and the $F-P$ phase transitions.

\textbf{II.4 Power Law Relations and the Infinite Lattice Statical Critical
Exponents}

On the other hand, the critical exponent values for $df-F$\ phase transition
can be obtained using the following power law relations $\left[ 53\right] $.

\begin{equation}
M(L)=\varepsilon ^{\beta (L)}
\end{equation}

\begin{equation}
\chi (L)=\varepsilon ^{-\gamma (L)}
\end{equation}

\begin{equation}
C(L)=\varepsilon ^{-\alpha (L)}
\end{equation}%
where $\varepsilon =$($T-T_{C}(L)$)$/T_{C}(L)$. The finite lattice critical
exponents $\beta (L)$, $\beta ^{\prime }(L)$, $\gamma (L)$, $\gamma ^{\prime
}(L)$, $\alpha (L)$ and $\alpha ^{\prime }(L)$ of the order parameter ($M$),
susceptibility ($\chi $) and the specific heat ($C$) quantities are obtained
from the slope of the log-log plot of the power laws relations for each
finite lattices in the interval $0.05\leq \varepsilon \leq 0.2$. The
infinite lattice critical exponents are obtained using linear extrapolation
and their values are given in Table I. The estimated values for cooling
algorithm and three simulation sets of heating algorithm are in good
agreement with the finite size scaling critical exponent estimations and the
universal values for $3d$ Ising model ($\beta =0.31$, $\gamma =1.25$, $%
\alpha =\alpha ^{\prime }=0.12$ and $\nu =0.64$).

\begin{center}
{\small Table 1. The estimated values of the infinite lattice critical
exponents and the critical temperatures (}$\alpha ${\small , }$\beta $%
{\small , }$\gamma ${\small \ and }$T_{C1}^{\chi ,C}${\small (}$\infty $%
{\small )) using linear extrapolation.}

\begin{tabular}{||c|c|c|c|c|c||}
\hline\hline
\multicolumn{6}{||c||}{$df$\textbf{\ - }$F$\textbf{\ phase transition}} \\ 
\hline\hline
& \multicolumn{4}{|c|}{\textbf{Heating}} & \textbf{Cooling} \\ \hline
& \textbf{Set I} & \textbf{Set II} & \textbf{Set III} & \textbf{Average of
sets} &  \\ \hline
$T_{C1}^{\chi }\mathbf{(}\infty \mathbf{)}$ & $1.50\pm 0.02$ & $1.50\pm 0.03$
& $1.52\pm 0.03$ & $1.51\pm 0.03$ & $1.52\pm 0.04$ \\ \hline
$T_{C1}^{C}\mathbf{(}\infty \mathbf{)}$ & $-$ & $-$ & $1.52\pm 0.01$ & $-$ & 
$1.52\pm 0.01$ \\ \hline
$\mathbf{\beta (T<T}_{\mathbf{C1}}\mathbf{)}$ & $0.31\pm 0.01$ & $0.31\pm
0.01$ & $0.30\pm 0.01$ & $0.31\pm 0.01$ & $0.31\pm 0.01$ \\ \hline
$\mathbf{\gamma (T<T}_{\mathbf{C1}}\mathbf{)}$ & $1.25\pm 0.01$ & $1.23\pm
0.03$ & $1.23\pm 0.02$ & $1.24\pm 0.03$ & $1.23\pm 0.02$ \\ \hline
$\mathbf{\alpha }\mathbf{(T<T}_{\mathbf{C1}}\mathbf{)}$ & $0.12\pm 0.01$ & $%
0.12\pm 0.01$ & $0.11\pm 0.01$ & $0.12\pm 0.01$ & $0.12\pm 0.01$ \\ \hline
$\mathbf{\alpha }^{\prime }\mathbf{(T>T}_{\mathbf{C1}}\mathbf{)}$ & $0.12\pm
0.01$ & $0.12\pm 0.01$ & $0.12\pm 0.01$ & $0.12\pm 0.01$ & $0.12\pm 0.01$ \\ 
\hline\hline
\end{tabular}
\end{center}

\bigskip

\textbf{III. SUMMARY}

The ($kT_{C}/J$, $d$) phase diagrams is illustrated in figure 9 for the
presence of $df$ order. The type of special point is determined by the $d=2.9
$ parameter. The calculations show that model exhibits the phase transition
from order to order for $d=2.9$ and the first order phase transition from
order to disorder in the $3\leq d<4$ parameter region. If the model doesn't
exhibit the $df$ ordered phase instead of the $F$ ordered phase in the low
temperature region, the continuous phase transition from $F$ order to $F$
order ($F-F$) is considered as the weak first order. This constitutes the
part of the first order phase transition line which creates the critical end
point ($CEP$) $\left[ 7,8\right] $. However, CA results show that, the model
has a $df$ ordered phase for the parameters in the $2.9\leq d<4.0$ region.
The spin system contains $S=+1$ and $S=0$ states. As a result of this, the
order parameters $M$ and $Q$ are almost equal each other ($M\cong Q\neq 0)$
at low temperatures. With increasing temperature, the dense ferromagnetic ($%
df$) ordered phase changes continuously to ferromagnetic ($F$) ordered phase
at the $d=2.9$, $k=-0.5$ with the enough contribution of $S=-1$ state.
Therefore, near the $d=2.9$, the first order phase transition line have been
changed to second order phase transition line, and there occurs the
tricritical point $TCP$ (Figure 9). In order to determine the universality
class of the successive $df-F-P$ second order phase transitions, the static
critical exponents ( $\alpha $, $\beta $, $\gamma $ and $\nu $) are
estimated within the framework of the finite - size scaling theory. The
estimated values of the critical exponents ( $\alpha =0.12$, $\beta =0.31$, $%
\gamma =1.25$ and $\nu =0.64$) near the $T_{C1}$ and $T_{C2}$ temperatures
are in good agreement with the theoretical values for three sets. The $df-F$
phase transition is analyzed with the power laws for comparing with the
critical exponent values estimated from the finite - size scaling theory.
The obtained values are in compatible with the finite - size scaling analyze
results and the universal values for the $3d$ Ising model. The obtained
results have shown that the $df-F$ phase transition is of the second order
and it is compatible with the universal Ising critical behavior for $d=2.9$
parameter value through $k=-0.5$ line. As a result of this, the definition
of the $df$ phase changes the phase transition type and the special point
type in the phase space for the BEG model. This result will lead to
reexamine the structure of phase spaces.

\textbf{ACKNOWLEDGEMENT}

This work is supported by a grant from Gazi University (BAP:05/2003-07).

\textbf{References}

$\left[ 1\right] $ Keskin M, Ekiz C, Yal\c{c}\i n O, 1999 \textit{Physica A} 
\textbf{267} 392

$\left[ 2\right] $ Albayrak E, Keskin M, 2000 \textit{J. Magn. Magn. Mater.} 
\textbf{203} 201 

$\left[ 3\right] $ Ekiz C, Keskin M, 2002 \textit{Phys. Rev. B} \textbf{66}
054105

$\left[ 4\right] $ \"{O}zkan A, Kutlu B, 2010 \textit{Int. J. of Mod. Phys.
B }accepted to publish

$\left[ 5\right] $ \ Sefero\u{g}lu N, Kutlu B, 2007 \textit{Physica A} 
\textbf{374} 165 

$\left[ 6\right] $\ \"{O}zkan A, Kutlu B, 2007 \textit{Int. J. of Mod. Phys.
C} \textbf{18} 1417

$\left[ 7\right] $ Hoston W, Berker A N, 1991 \textit{Phys. Rev. Lett}. 
\textbf{67} 1027 

$\left[ 8\right] $ Netz R R, Berker A N, 1993 \textit{Phys. Rev. B} \textbf{%
47} 15019

$\left[ 9\right] $ Netz R R, 1992 \textit{Europhys. Lett.} \textbf{17} 373 

$\left[ 10\right] $ Baran O R, Levitskii R R, 2002 \textit{Phys. Rev. B} 
\textbf{65} 172407

$\left[ 11\right] $ Lapinskas S, Rosengren A, 1993 \textit{Phys. Rev. B} 
\textbf{49} 15190

$\left[ 12\right] $ Blume M, Emery V J and Griffiths R B, 1971 \textit{Phys.
Rev. A} \textbf{4} 1071 

$\left[ 13\right] $ Lajzerowicz J and Siverdi\.{e}re J, 1975 \textit{Phys.
Rev. A} \textbf{11} 2090 

$\left[ 14\right] $ Lajzerowicz J and Siverdi\.{e}re J, 1975 \textit{Phys.
Rev}. A \textbf{11} 2101 

$\left[ 15\right] $ Schick M and Shih W H, 1986 \textit{Phys. Rev. B} 
\textbf{34} 1797 

$\left[ 16\right] $ Newman K E and Dow J D, 1983 \textit{Phys. Rev. B} 
\textbf{27} 7495 

$\left[ 17\right] $ Gu B L, Newman K E, Fedders P A, 1987 \textit{Phys. Rev.
B} \textbf{35} 9135 

$\left[ 18\right] $ Gu B L, Ni J, Zhu J L, 1992 \textit{Phys. Rev. B} 
\textbf{45} 4071 

$\left[ 19\right] $ Lawrie I D, Sarbach S, \textit{Phase transitions and
Critical Phenomena, edited by C. Domb and J. L. Lebowitz 1984 Vol \textbf{9}
Academic Press, New York \qquad\ }\ \ \ \ 

$\left[ 20\right] $ Kessler M, Dieterich W and Majhofer A, 2003 \textit{%
Phys. Rev. B} \textbf{\ 67} 134201 

$\left[ 21\right] $ Ausloos M, Clippe P, Kowalski J M, Pekalski A, 1980 
\textit{Phys. \ Rev. A} \textbf{22}\textit{\ }2218, \textit{ibid}. 1980 
\textit{IEEE Trans. Magnetica MAG }\textbf{16} 233\textit{\  }

$\left[ 22\right] $ Ausloos M, Clippe P, Kowalski J M, Pekalska J, Pekalski
A,\textit{\ }1983\textit{\ Phys. Rev. A} \textbf{28}\textit{\ 3080}; Droz M,
Ausloos M, Gunton J D, \textit{ibid.}\ 1978 \textbf{18}\textit{\ 388 } 

$\left[ 23\right] $ Ausloos M, Clippe P, Kowalski J M, Ekalska J P, Pekalski
A, 1983 \textit{J. Magnet. and Magnet. Matter} \textbf{39} 21 

$\left[ 24\right] $ Koza Z, Jasuukiewicz C, Pekalski A, 1990 \textit{Physica
A } \textbf{\ 164 } 191 

$\left[ 25\right] $ Saul D M, Wortis M and Stauffer D, 1974 \textit{Phys.
Rev.} B \textbf{9} 4964

$\left[ 26\right] $ Takahashi K, Tanaka M, 1979 \textit{J. Phys. Soc. Japan} 
\textbf{46} 1428 

$\left[ 27\right] $ Berker A N, Wortis M, 1976 \textit{Phys. Rev. B} \textbf{%
14 } 4946

$\left[ 28\right] $ \"{O}zkan A, Kutlu B, 2009 \textit{Int. J. of Mod. Phys.
C }\textbf{20}\textit{\ }1617

$\left[ 29\right] $ \"{O}zkan A, Kutlu B, 2010 \textit{Cent. Europ. J. of
Phys.}, DOI: 10.2478/S11534-010-0081-1 

$\left[ 30\right] $ Demirel H, \"{O}zkan A, Kutlu B, 2008 \textit{Chineese
Phys. Lett. }\textbf{25}\textit{\ 2599}

$\left[ 31\right] $ Kutlu B, 2001 \textit{Int. J. Mod. Phys. C} \textbf{12}
1401 

$\left[ 32\right] $ Kutlu B, 2003 \textit{Int. J. Mod. Phys. C} \textbf{14}
1305 

$\left[ 33\right] $ Solak A and Kutlu B, 2004 \textit{Int. J. Mod. Phys. C} 
\textbf{15} 1425 

$\left[ 34\right] $ Kutlu B, \"{O}zkan A, Sefero\u{g}lu N, Solak A and Binal
B, 2005 \textit{Int. J. Mod. Phys. C}\textbf{\ 16} 933

$\left[ 35\right] $ \"{O}zkan A, Sefero\u{g}lu N and Kutlu B, \textit{\ 2006
Physica A} \textbf{362} 327 

$\left[ 36\right] $ Sefero\u{g}lu N, \"{O}zkan A and Kutlu B, \textit{\ 2006
Chineese Phys. Lett.} \textbf{23} 2526

$\left[ 37\right] $ Kutlu B, Aktekin N, 1994 \textit{J. Stat. Phys.} \textbf{%
75} 757 

$\left[ 38\right] $ Kutlu B, Aktekin N, 1995 \textit{Physica A} \textbf{215}
370 

$\left[ 39\right] $ Kutlu B, 1997 \textit{Physica A} \textbf{234} 807 

$\left[ 40\right] $ Kutlu B, 1997 \textit{Physica A} \textbf{243} 199

$\left[ 41\right] $ Aktekin N, \ 2000 \textit{Annual Reviews of
Computational Physics}\textbf{\ VII }, ed. D.Stauffer, pp.1.World
Scientific, Singapore

$\left[ 42\right] $ Aktekin N, 2001 \textit{J. stat. Phys. }\textbf{104}
1397 

$\left[ 43\right] $ Aktekin N, Erko\c{c} S, 2001 \textit{Physica A} \textbf{%
290} 123 

$\left[ 44\right] $ Aktekin N, Erko\c{c} S, 2000 \textit{Physica A} \textbf{%
284} 206 

$\left[ 45\right] $ Merdan Z, Erdem R, 2004 \textit{Phys. Lett. A} \textbf{%
330} 403 

$\left[ 46\right] $ Merdan Z, Bay\i rl\i\ M, 2005 \textit{Applied
Mathematics and Computation} \textbf{167} 212 

$\left[ 47\right] $ Merdan Z, Atille D, 2007 \textit{Physica A} \textbf{376}
327 

$\left[ 48\right] $ Dress C, 1995 \textit{J. of physics A} \textbf{28} 7051 

$\left[ 49\right] $ Saito K, Takesue S and Miyashita S, 1999 \textit{Phys.
Rev. E} \textbf{59} 2783 

$\left[ 50\right] $ Kutlu B, \c{C}ivi M, 2006 \textit{Chineese Phys. Lett.} 
\textbf{23} 2670 

$\left[ 51\right] $ Creutz M, 1986 \textit{Ann. Phys.} \textbf{167} 62 

$\left[ 52\right] $ Binder K, 1981 \textit{Z}. \textit{Phys. B} \textbf{43}
119 

$\left[ 53\right] $ Huang K, 1987 \textit{Statistical Mechanics John Wiley
\& Sons }\textbf{396}

\bigskip 

\textbf{Figure Captions}

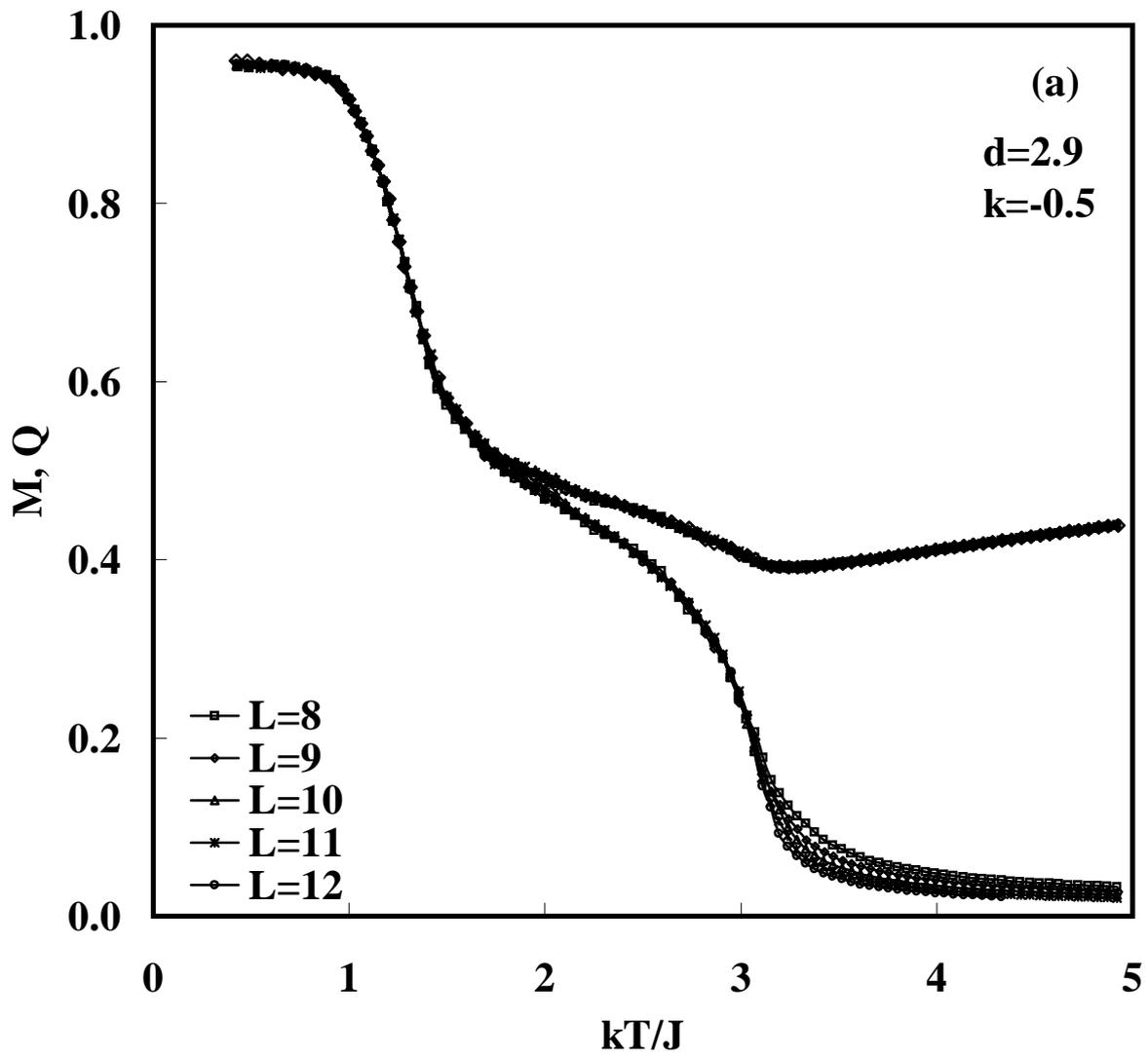
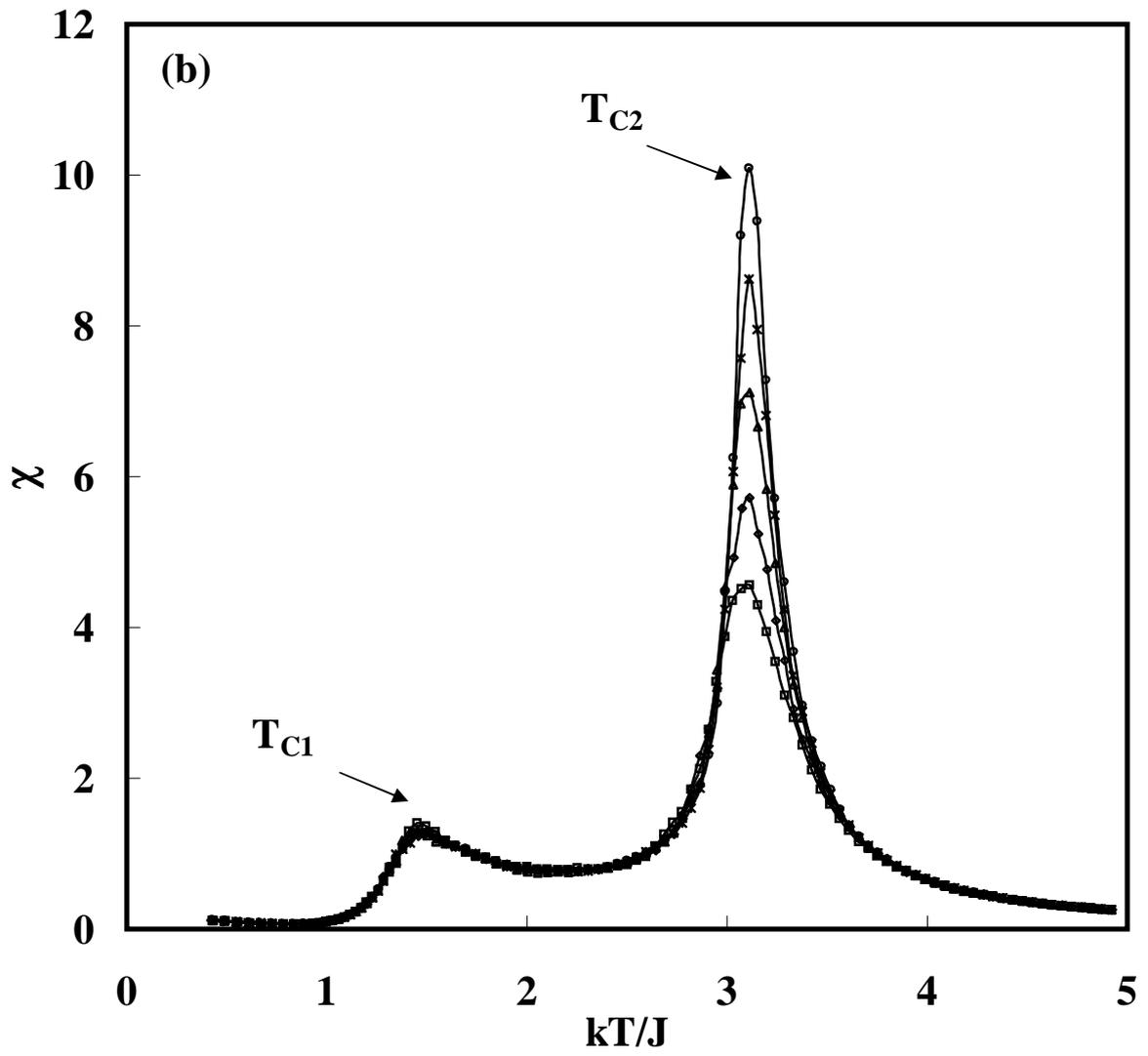
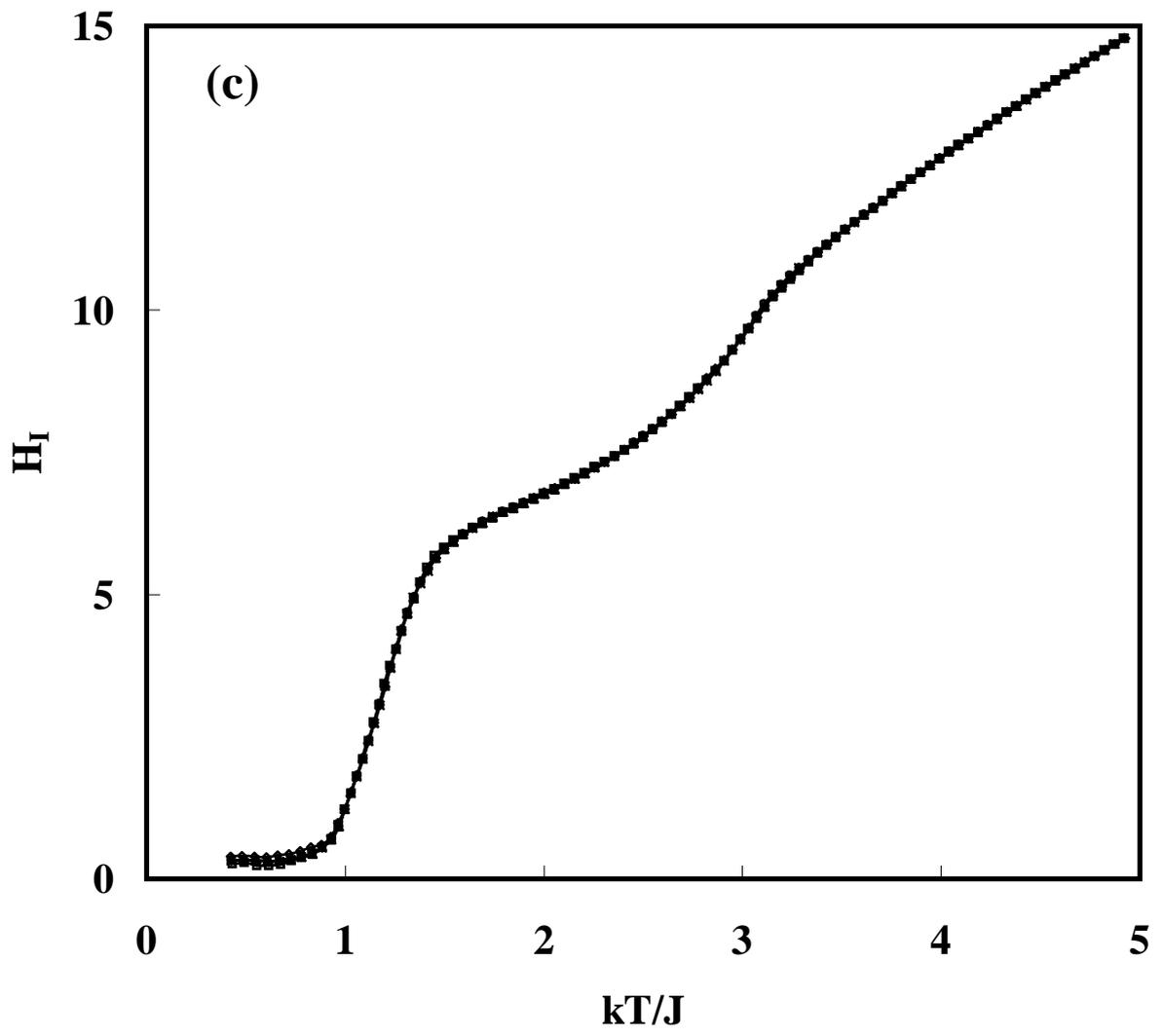
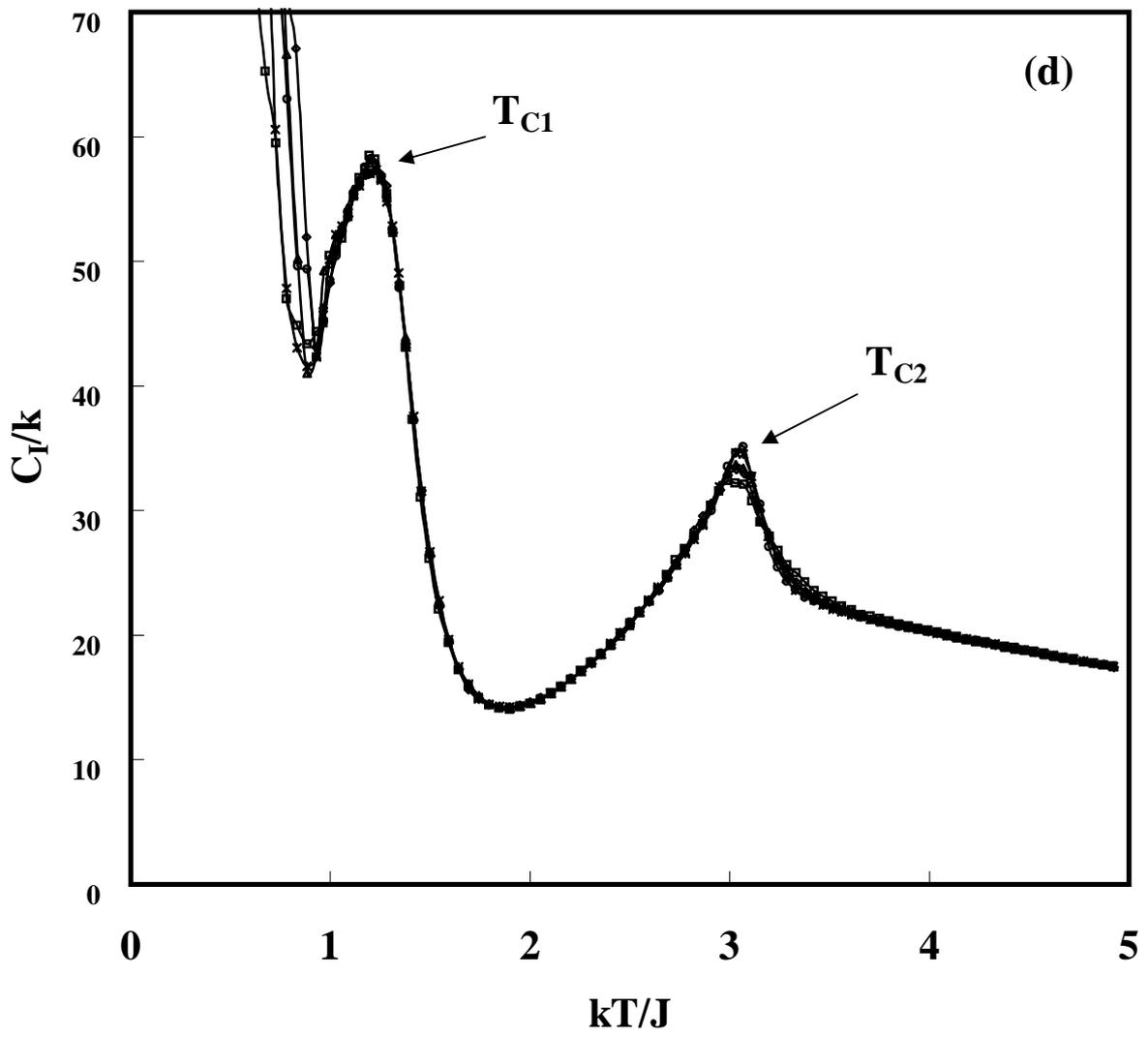
Figure 1. For ($d=2.9$, $k=-0.5$), the temperature dependence of (a) the
order parameters ($M$, $Q$), (b) the susceptibility ($\chi $), (c) the Ising
energy ($H_{I}$) and (d) the specific heat ($C_{I}/k$).

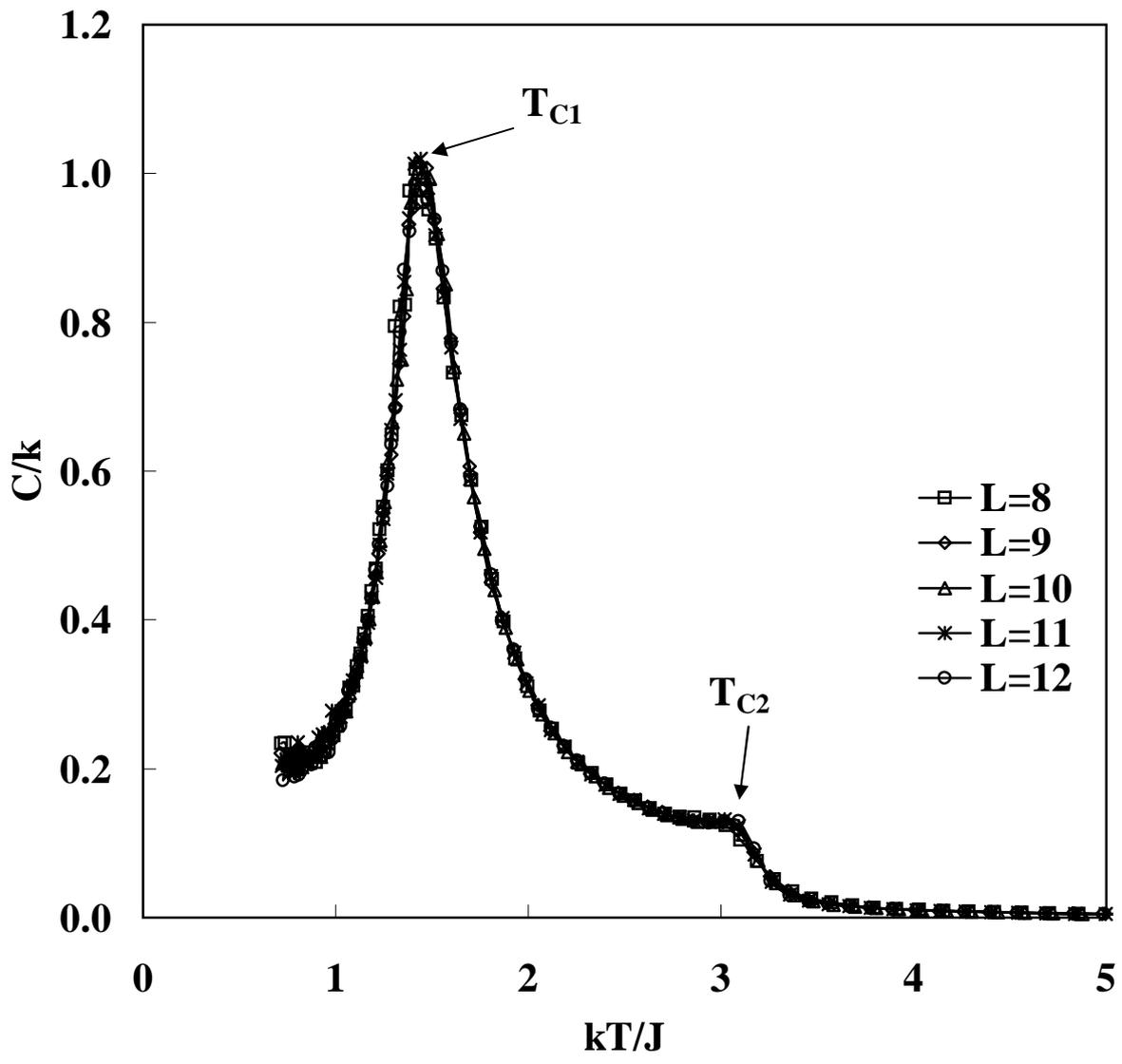
Figure 2. For ($d=2.9$, $k=-0.5$), the specific heat ($C/k$) calculated from 
$U$.

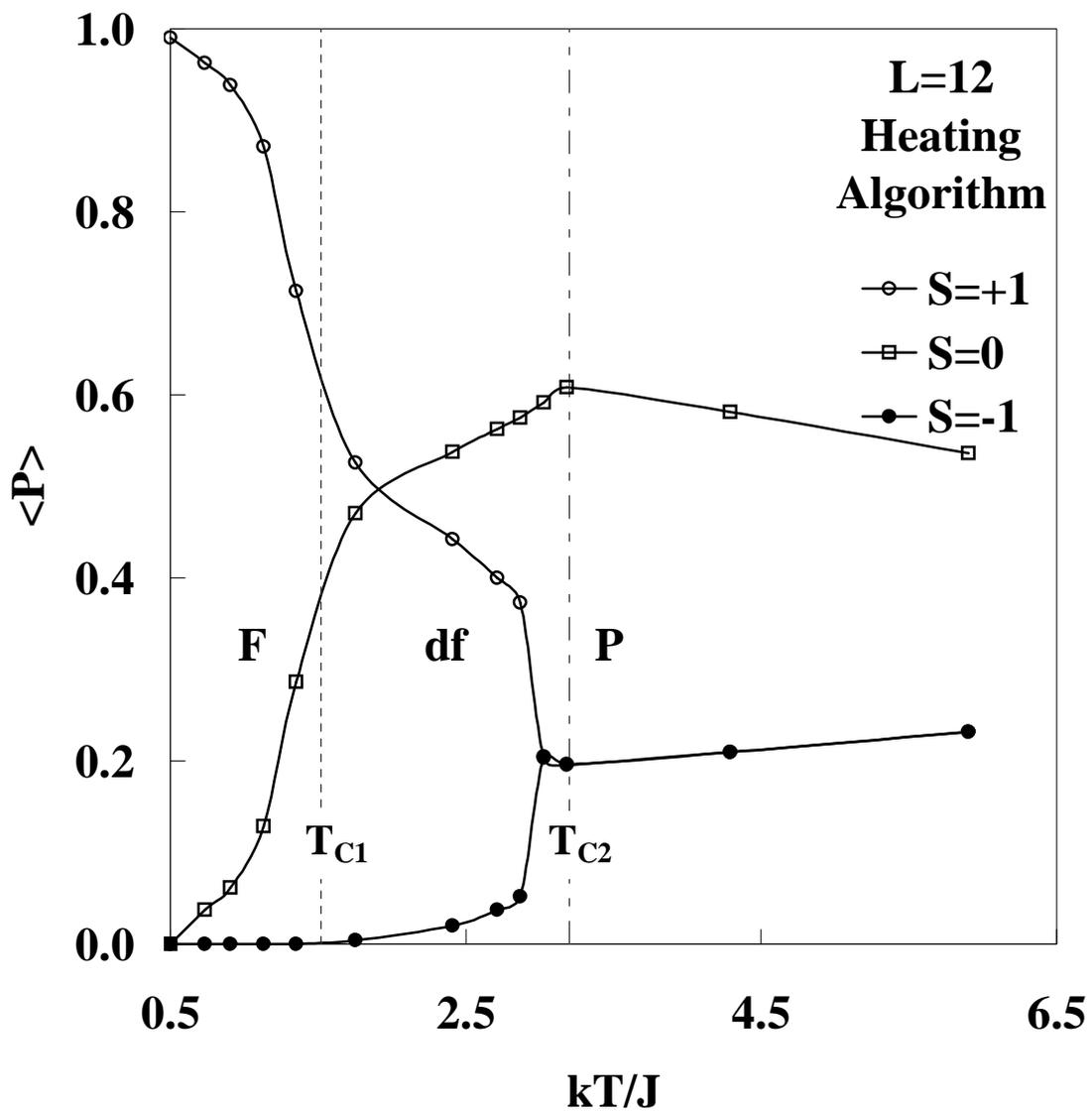
Figure 3. For ($d=2.9$, $k=-0.5$), the temperature dependences of the $%
\left\langle P\right\rangle $. $\left\langle P_{+1}\right\rangle $, \ $%
\left\langle P_{-1}\right\rangle $ and $\left\langle P_{0}\right\rangle $
correspond to the $S=+1$, $-1$ and $0$ spin states, respectively.

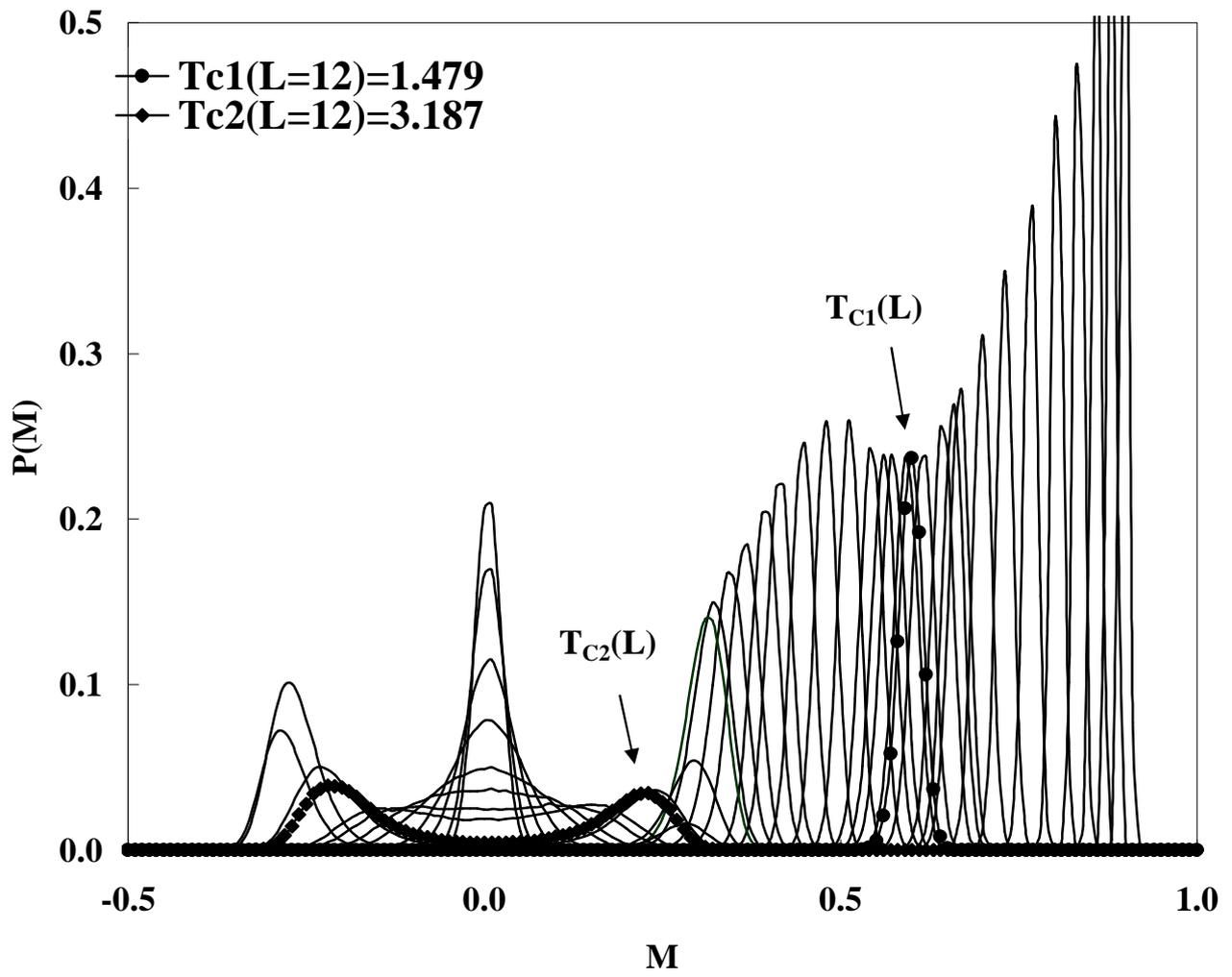
Figure 4. The probability distribution of the $M$ for ($d=-0.5$, $k=0.9$) on 
$L=12$.

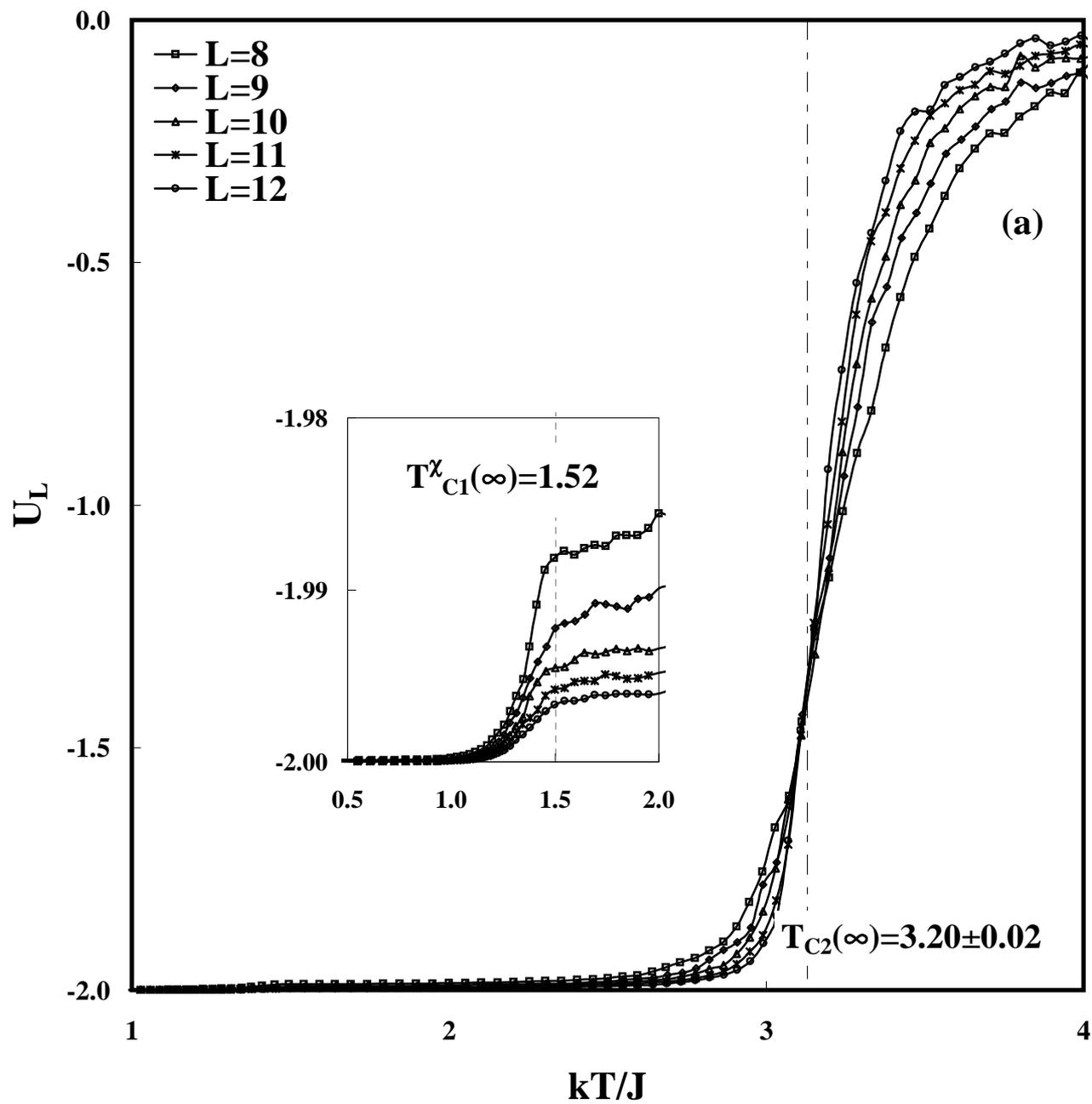
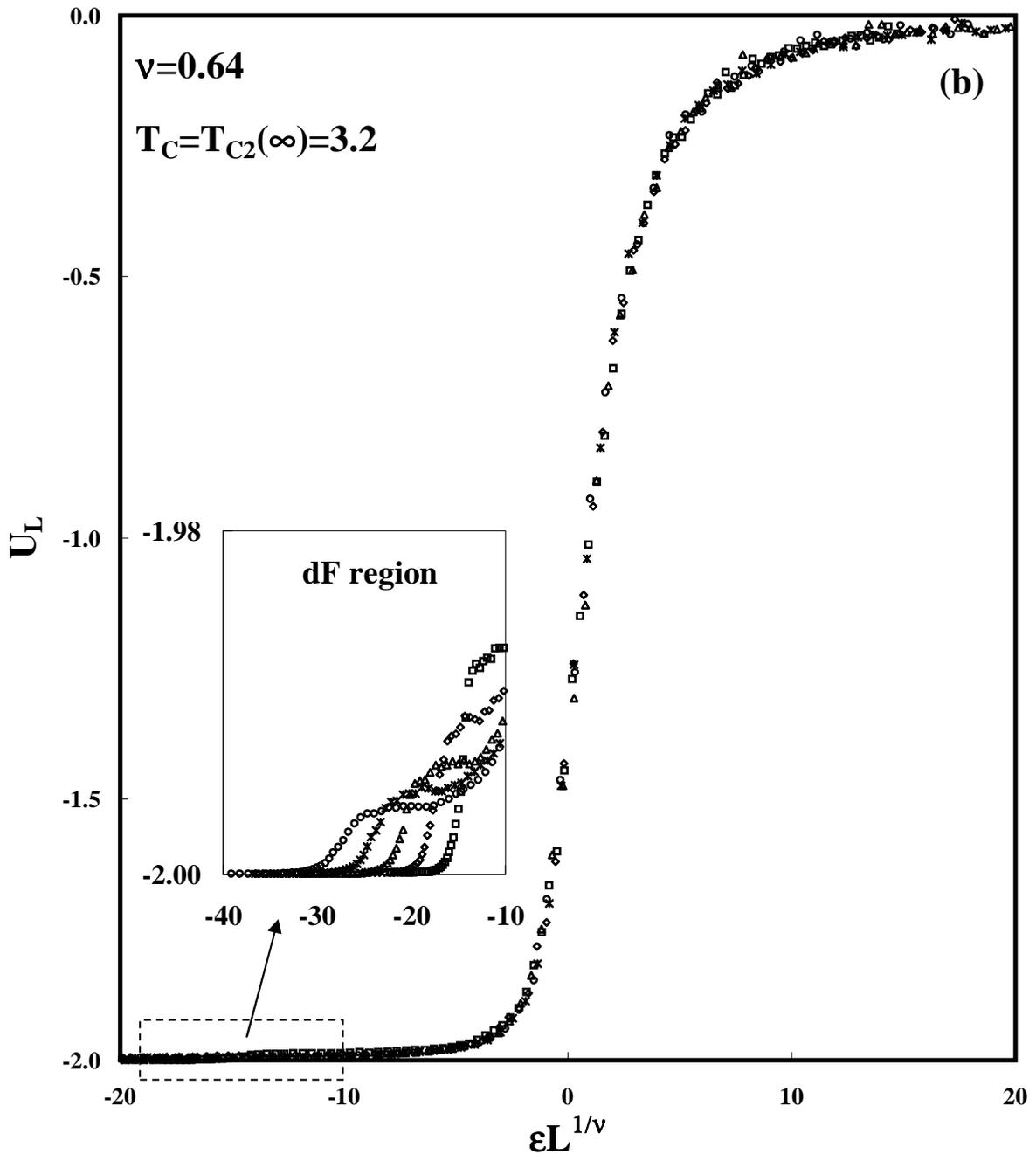
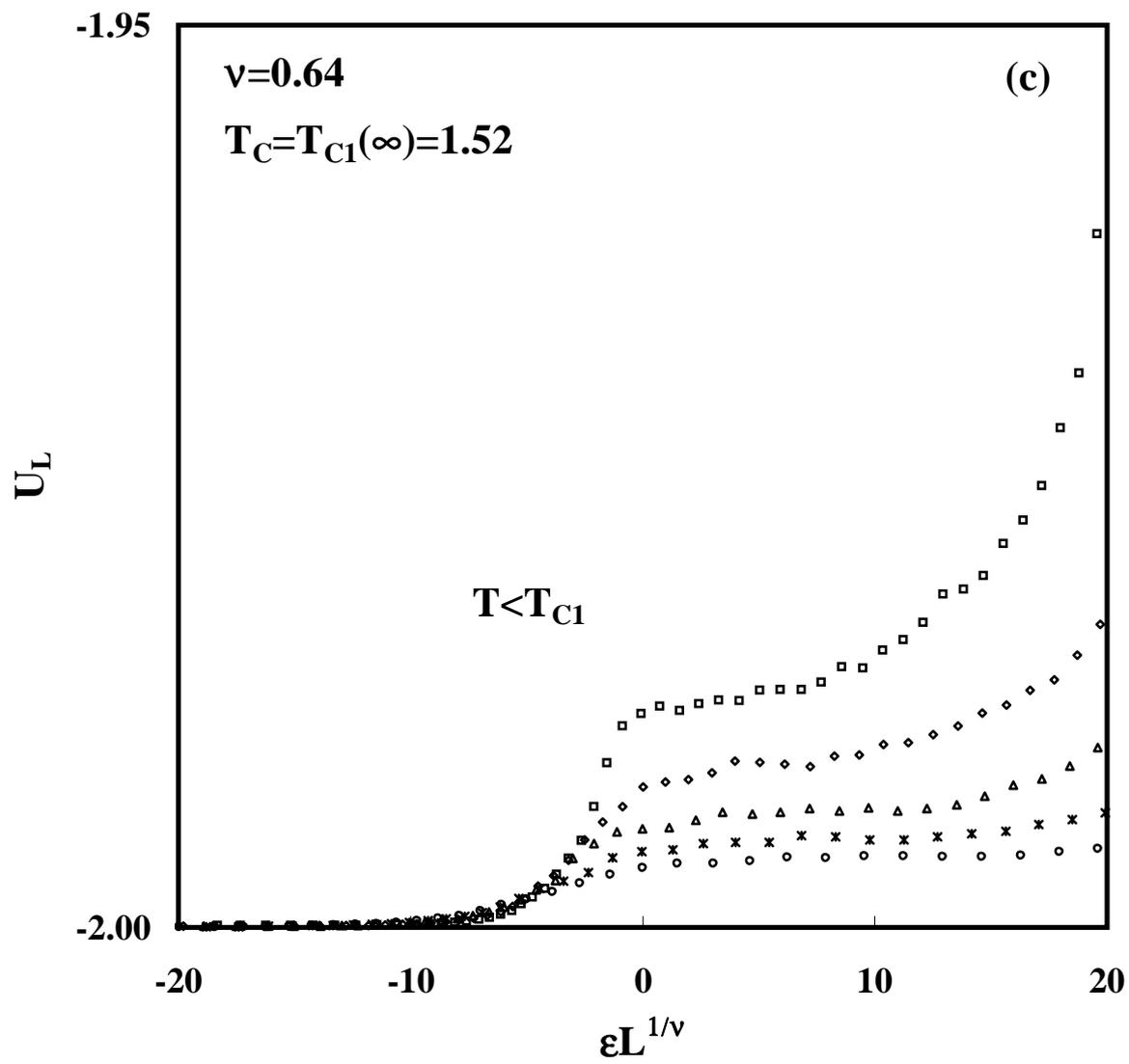
Figure 5. For ($d=2.9$, $k=-0.5$), (a) the temperature dependence of the
Binder cumulant ($U_{L}$), (b) the finite - size scaling of the Binder
cumulant near the $df$ $-$ $F$ $-$ $P$ phase transition with $%
T_{C2}^{U_{L}}(\infty )$, (c) the finite - size scaling of the Binder
cumulant near the $df$ $-$ $F$ phase transition with $T_{C1}^{{}}(\infty )$.

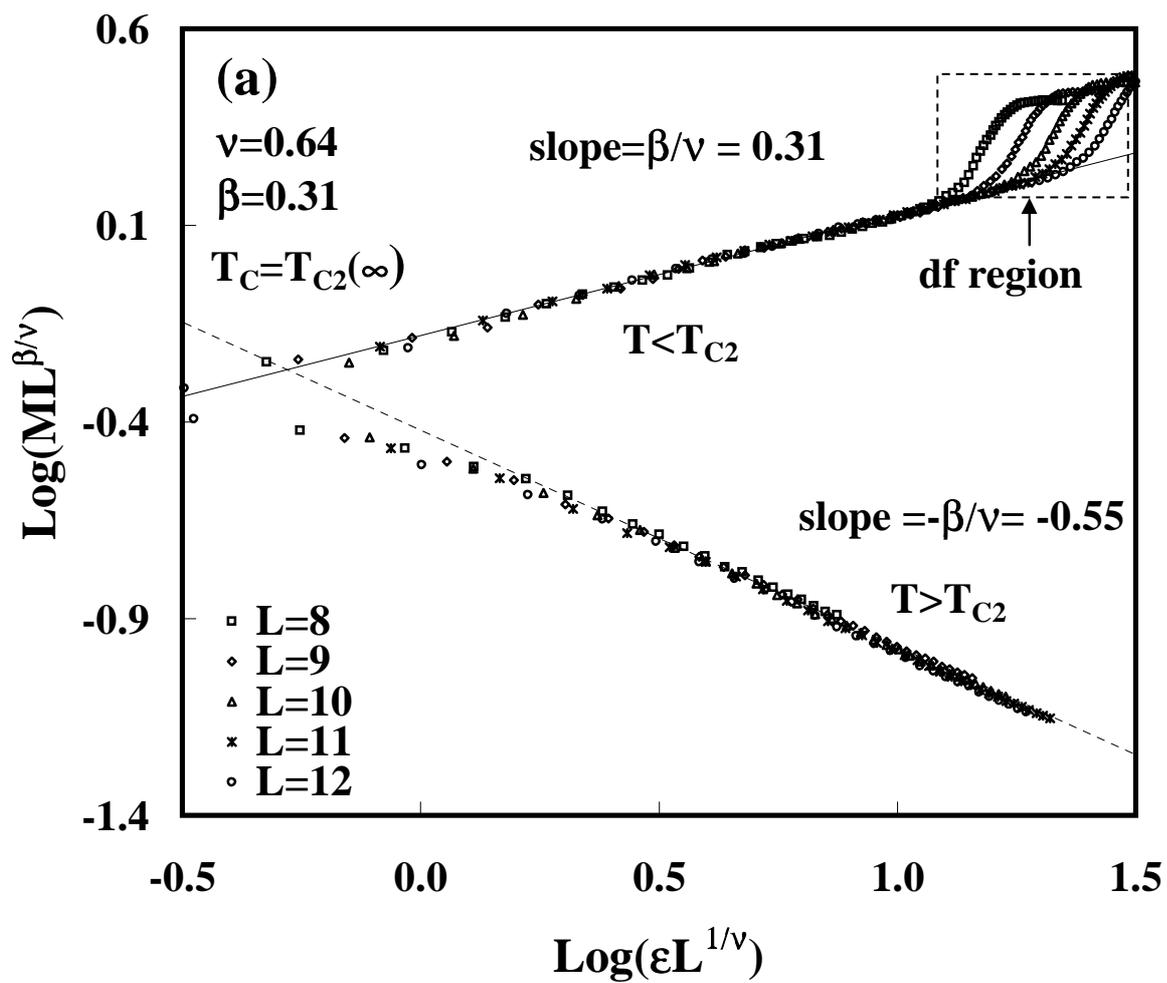
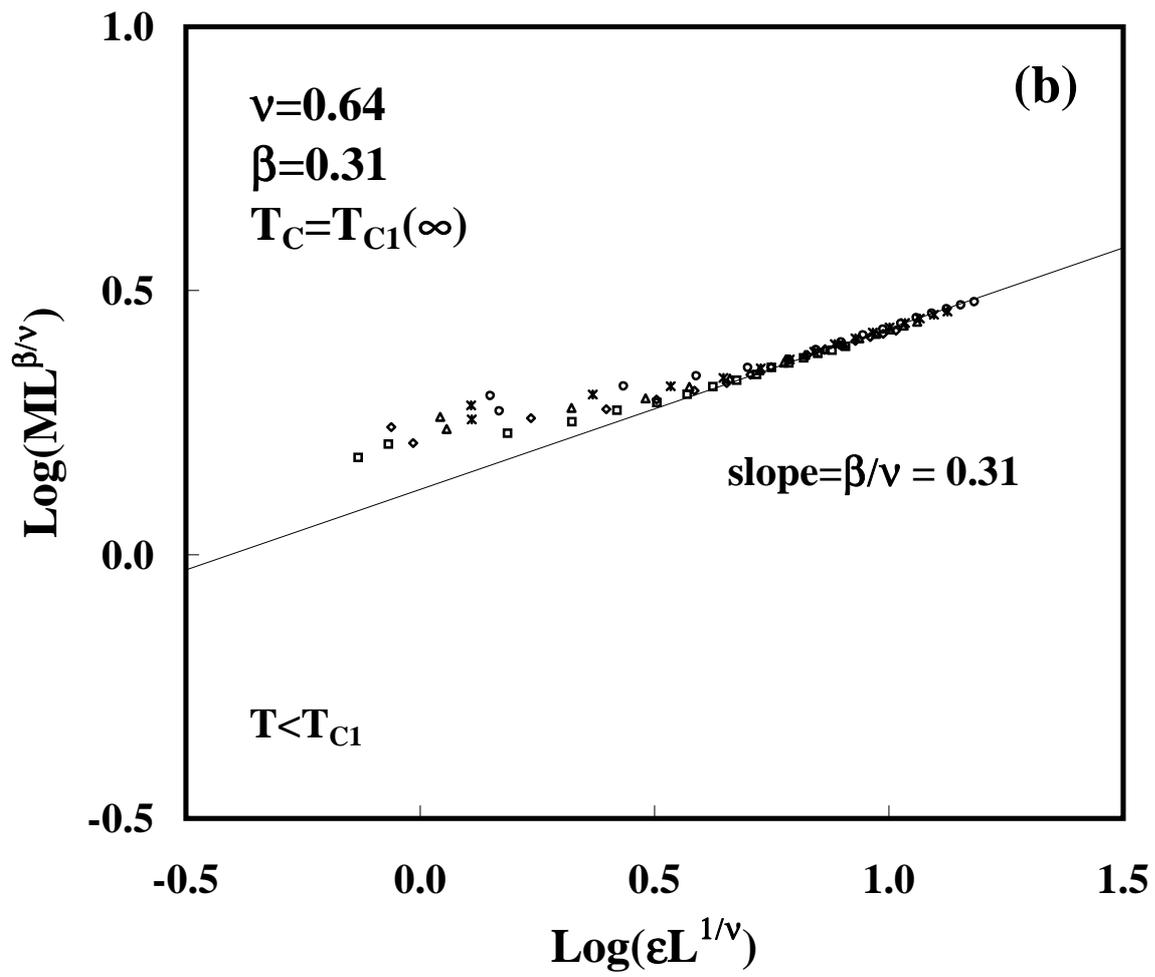
Figure 6. For ($d=2.9$, $k=-0.5$), the finite - size scaling plots of (a)
the order parameter with $T_{C2}^{{}}(\infty )$, (b) the order parameter
with $T_{C1}^{{}}(\infty )$ near the $df$\ $-$ $F$ phase transition for $%
T<T_{C1}^{{}}(\infty )$.

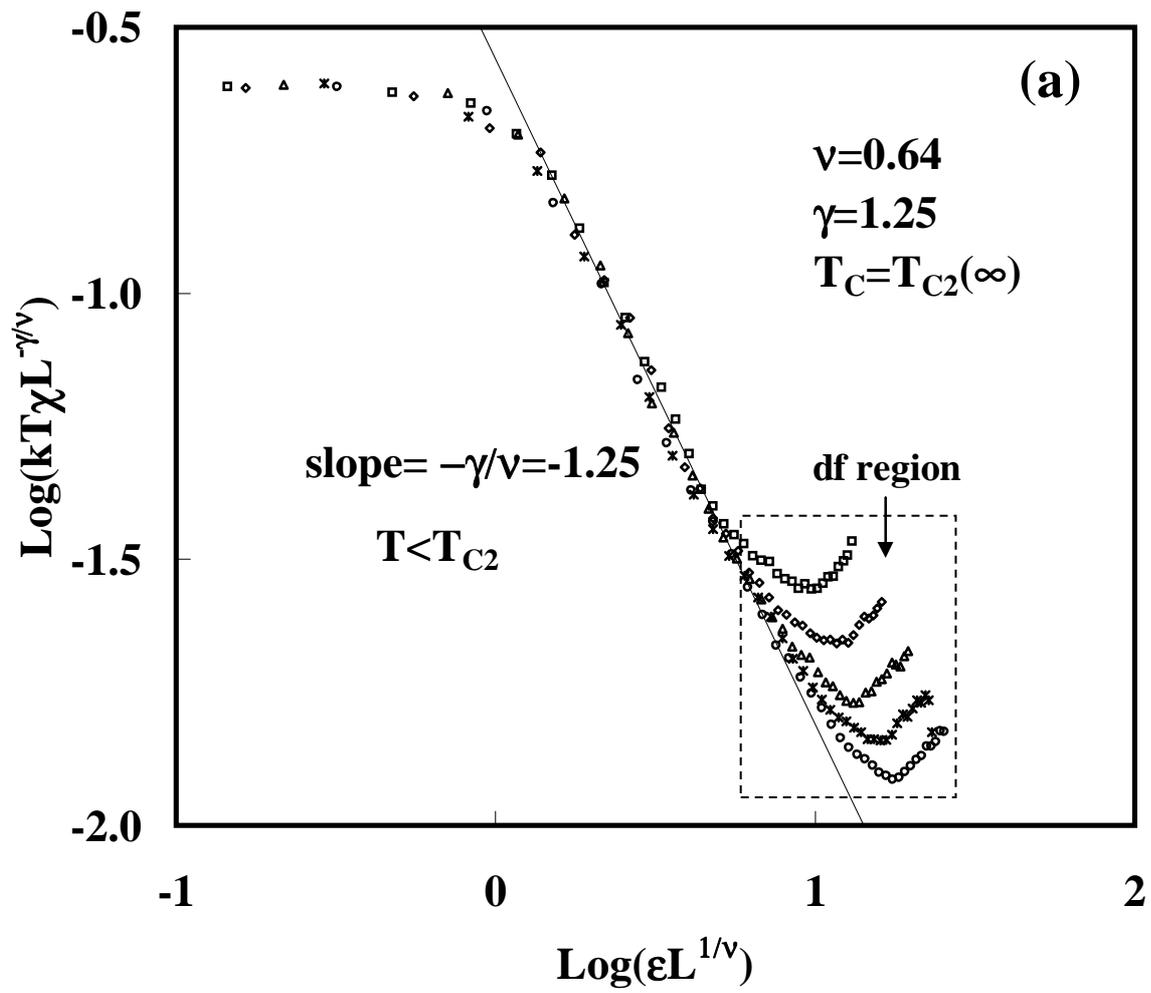
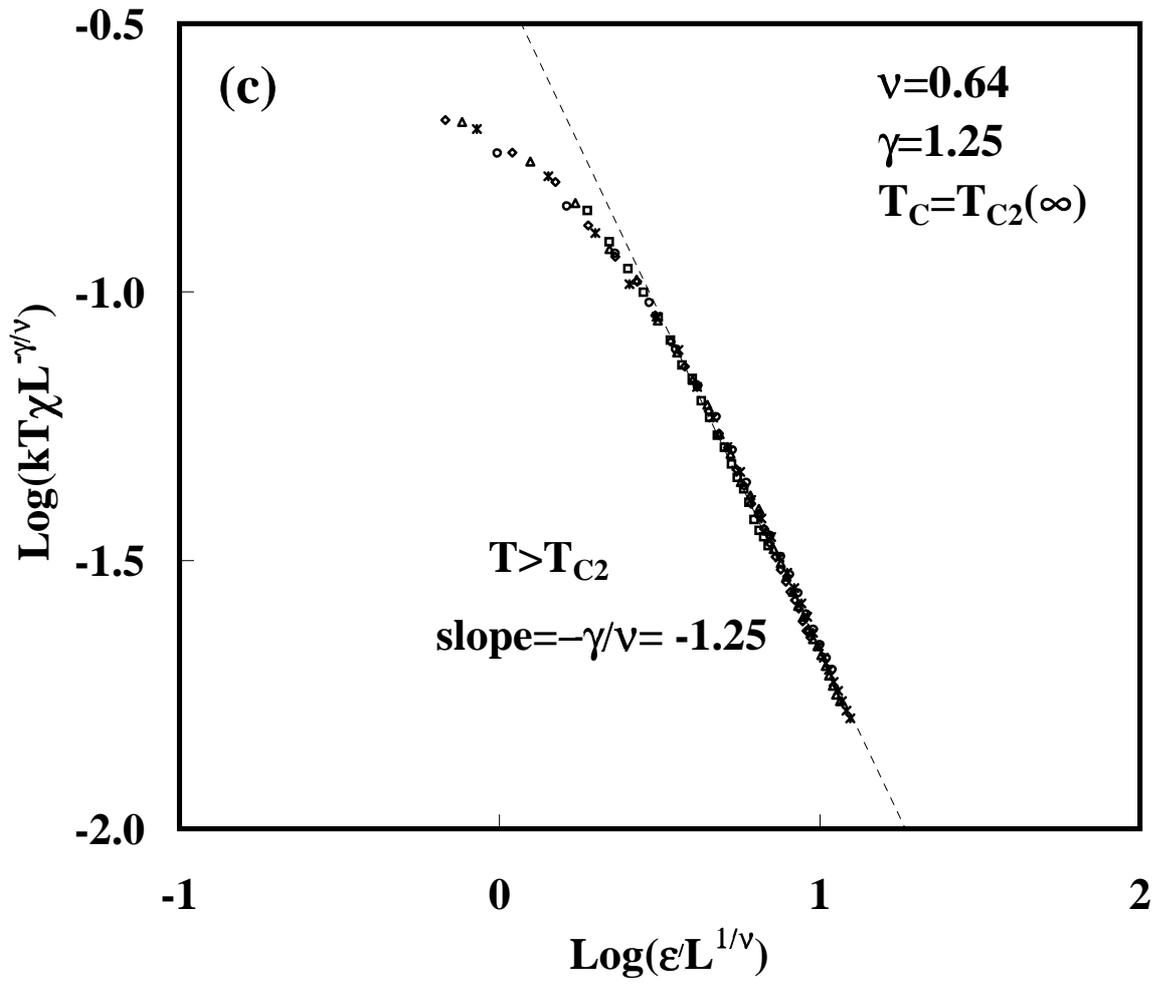
Figure 7. For ($d=2.9$, $k=-0.5$), the finite - size scaling plots of the
susceptibility (a) with $\varepsilon =$($T-T_{C2}^{{}}(\infty )$)/$%
T_{C2}^{{}}(\infty )$ for $T<T_{C2}^{{}}(\infty )$, (b) near the $df$ $-$ $F$
phase transition with $\varepsilon =$($T-T_{C1}^{{}}(\infty )$)/$%
T_{C1}^{{}}(\infty )$ for $T<T_{C1}^{{}}$($\infty $), (c) with $\varepsilon =
$($T-T_{C2}^{{}}(\infty )$)/$T$ for $T>T_{C2}^{{}}$($\infty $).

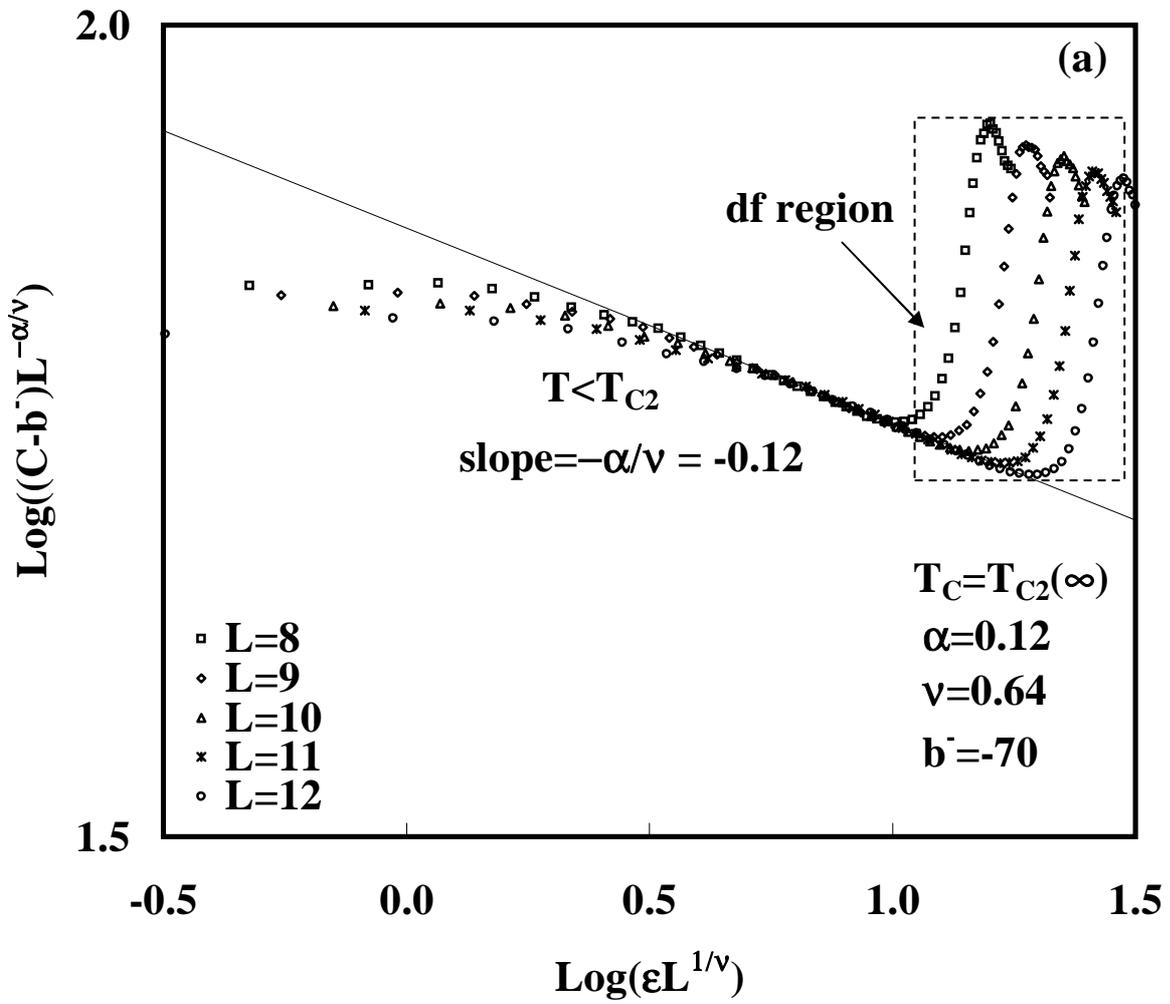
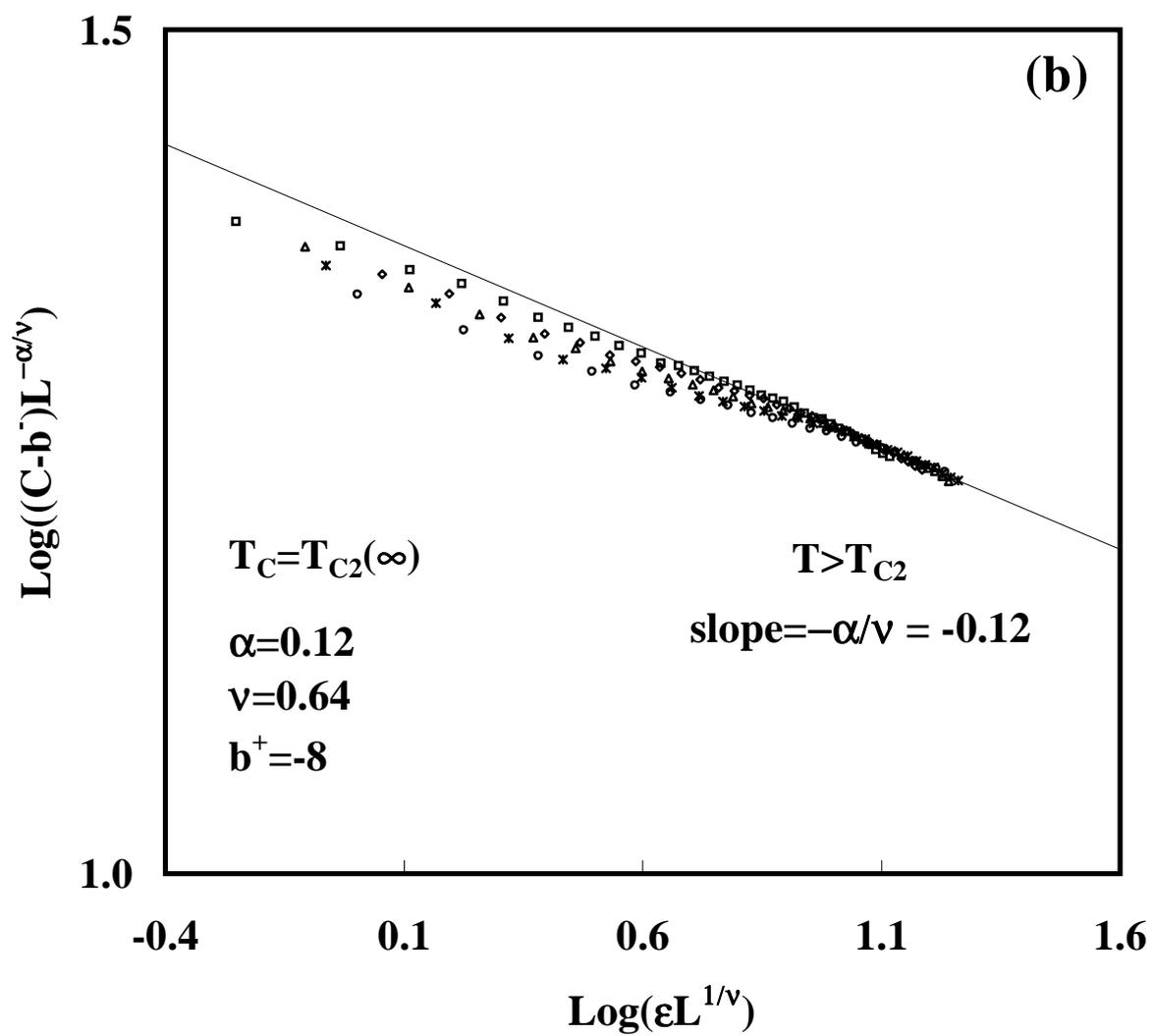
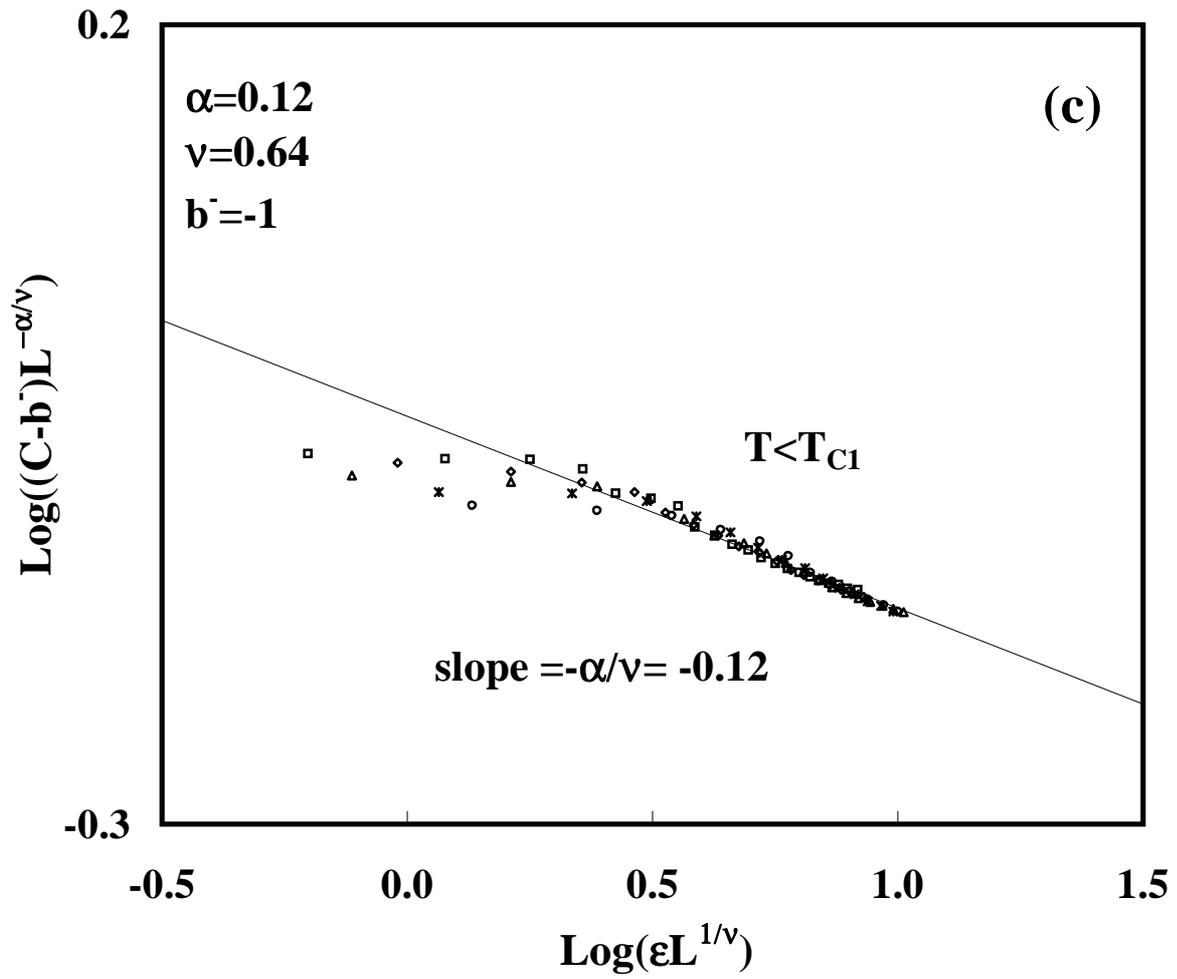
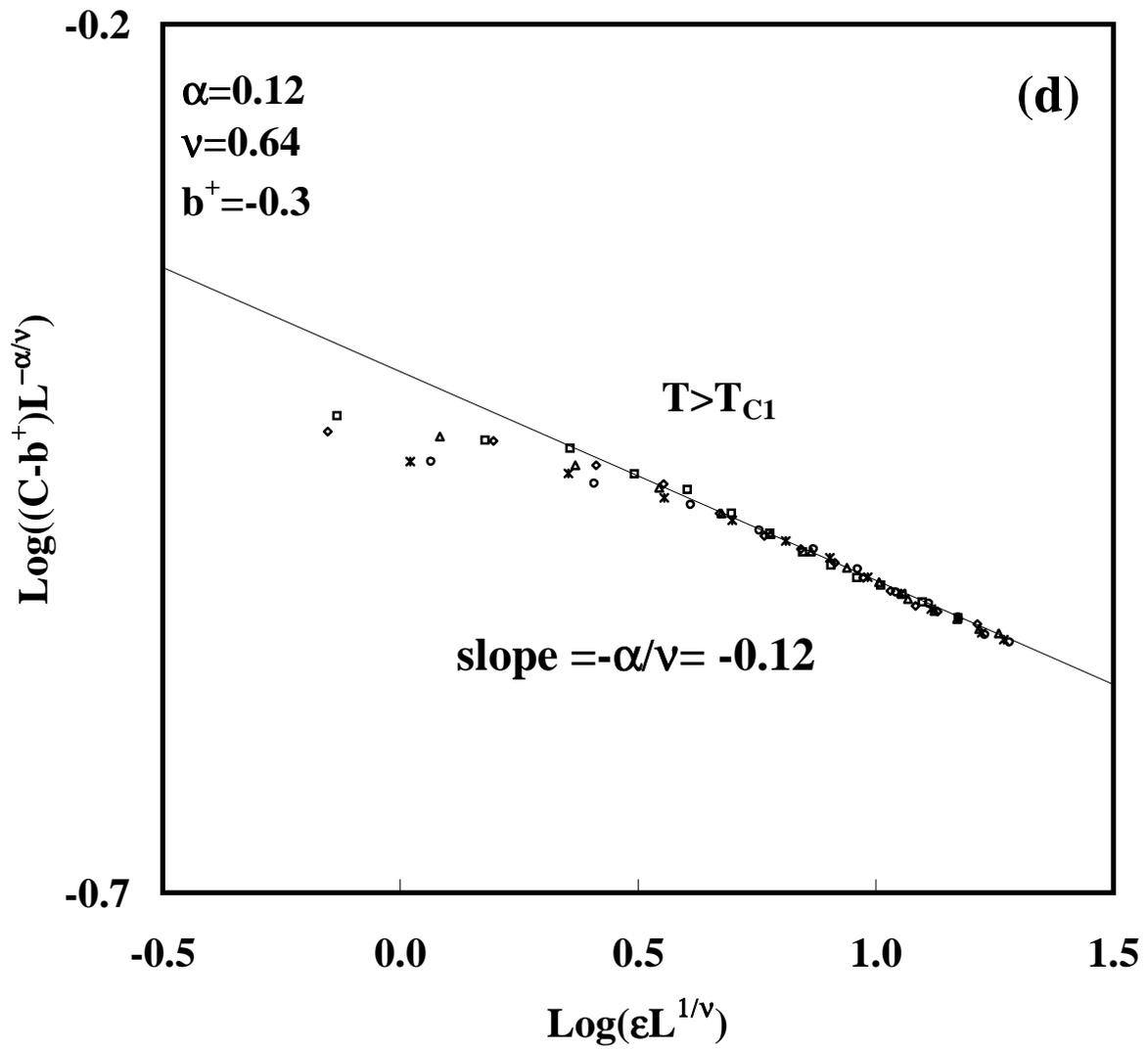
Figure 8. For ($d=2.9$, $k=-0.5$), the finite - size scaling plots of the
specific heat (a) for $T<T_{C2}^{{}}$($\infty $) with $T_{C2}^{{}}$($\infty $%
), (b) for $T>T_{C2}^{{}}(\infty )$ with $T_{C2}^{{}}(\infty )$, (c) for $%
T<T_{C1}^{C}$($\infty $) with $T_{C1}^{C}$($\infty $), (d) for $%
T>T_{C1}^{C}(\infty )$ with $T_{C1}^{C}(\infty )$.

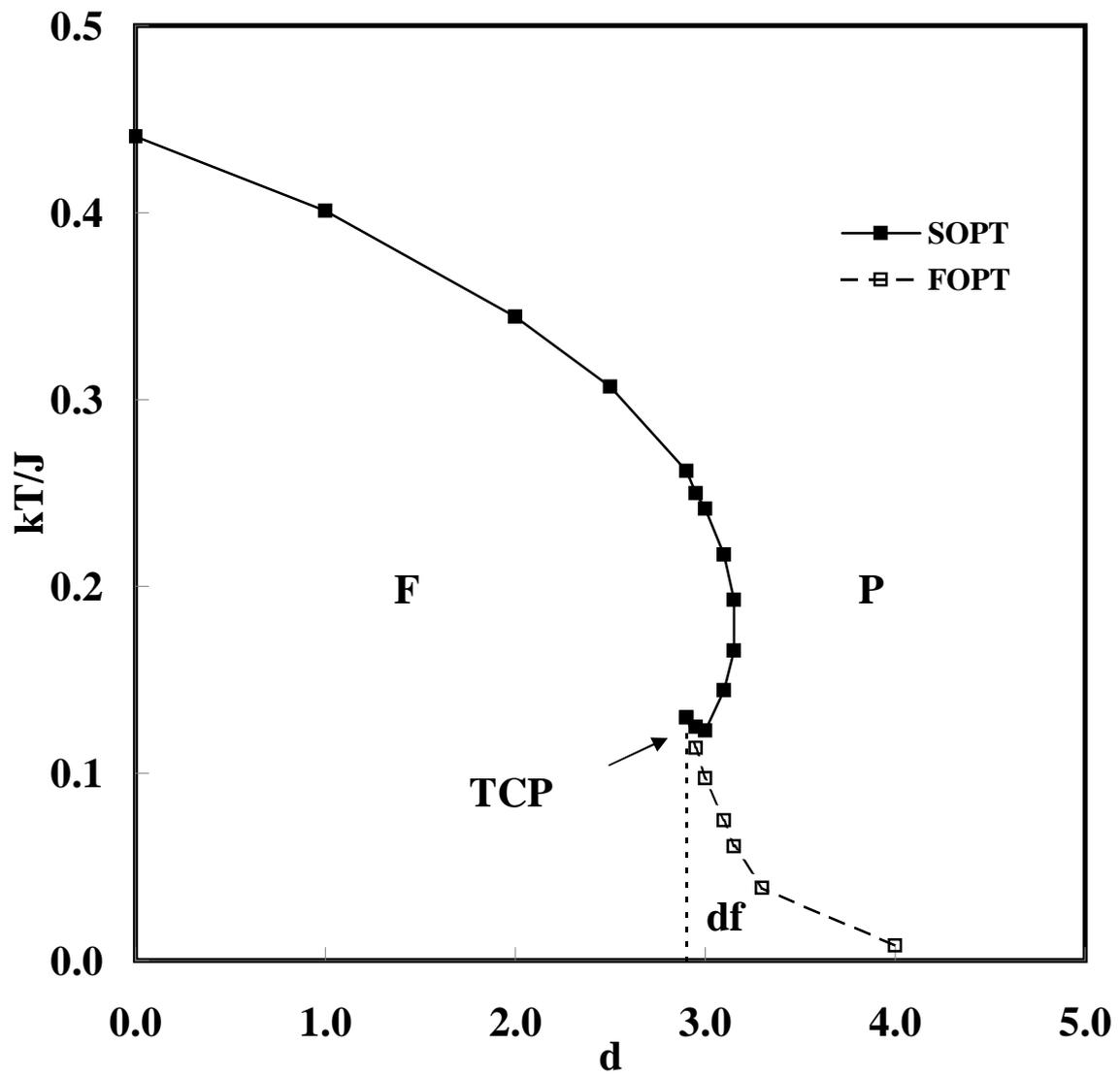
Figure 9. The phase diagram for $k=-0.5$. The phase space with $df$ phase
contains a $TCP$ at $d=2.9$.

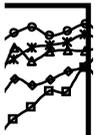
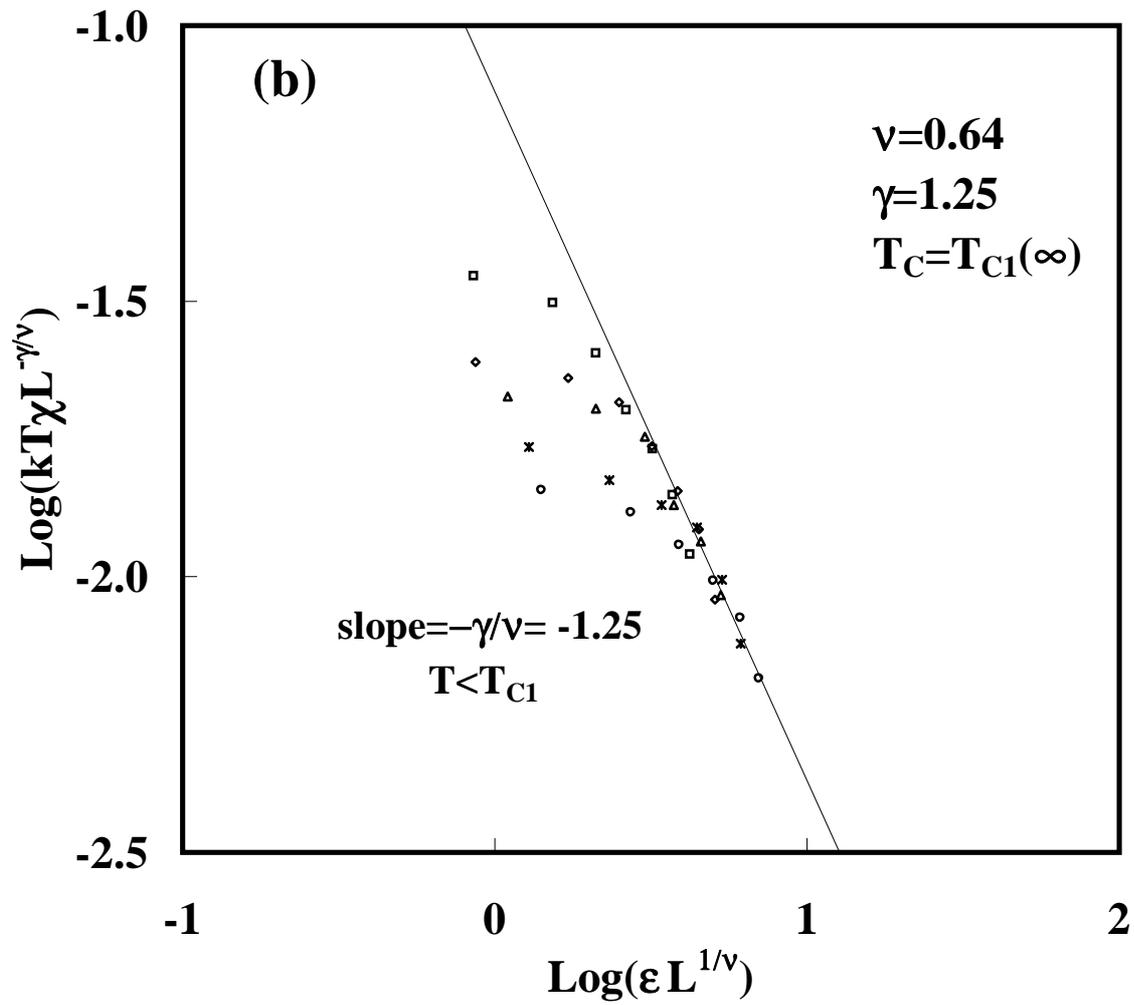
\end{document}